\documentclass[aip,graphicx]{revtex4-1}

\usepackage{amsmath}
\usepackage[english]{babel}
\usepackage{braket}
\usepackage{graphicx}
\usepackage{subcaption}
\usepackage{xcolor}
\usepackage{todonotes}
\usepackage[version=4]{mhchem}

\newcommand{\op}[1]{\ensuremath{\hat{#1}}}
\newcommand{\of}[1]{\ensuremath{\!\left(#1\right)}}
\newcommand{\myvec}[1]{\ensuremath{\mathbf{#1}}}
\newcommand{\mymat}[1]{\ensuremath{\mathbf{#1}}}

\newcommand{\ofX}[0]{\of{\myvec{X}}}
\newcommand{\ofzero}[0]{\of{\myvec{0}}}
\newcommand{\modeone}[0]{\sigma}
\newcommand{\modetwo}[0]{\tau}

\usepackage{newunicodechar}
\newunicodechar{→}{\ensuremath{\rightarrow}}

\begin{document}

\title{Coupling electrons and vibrations in molecular quantum chemistry}

\author{Thomas Dresselhaus}
\author{Callum B. A. Bungey}
\affiliation{Centre for Computational Chemistry, School of Chemistry, University of Bristol, Bristol BS8 1TS, United Kingdom}
\author{Peter J. Knowles}
\affiliation{School of Chemistry, Cardiff University, Main Building, Park Place, Cardiff CF10 3AT, United Kingdom}
\author{\newline Frederick R. Manby}
\email{fred.manby@bristol.ac.uk.}
\affiliation{Centre for Computational Chemistry, School of Chemistry, University of Bristol, Bristol BS8 1TS, United Kingdom}

\date{\today}

\begin{abstract}
We derive an electron-vibration model Hamiltonian in a quantum chemical framework, and explore the extent to which such a Hamiltonian can capture key effects of nonadiabatic dynamics. 
The model Hamiltonian is a simple two-body operator, and we make preliminary steps at applying standard quantum chemical methods to evaluating its properties, including mean-field theory, linear response, and a primitive correlated model. The Hamiltonian can be compared to standard vibronic Hamiltonians, but is constructed without reference to potential energy surfaces, through direct differentiation of the one- and two-electron integrals at a single reference geometry.
The nature of the model Hamiltonian in the harmonic and linear-coupling regime is investigated for pyrazine, where a 
simple time-dependent calculation including electron-vibration correlation is demonstrated to exhibit the well-studied population transfer between the S$_2$ and S$_1$ excited states.
\end{abstract}


\maketitle

\section{Introduction}
The Born--Oppenheimer approximation\cite{Born1927} is an indispensable framework for chemistry, providing the theoretical setting for the understanding of molecular structure; the ground-state potential-energy surface (PES) provides fundamental insight---through the work of Eyring and Polanyi \cite{Eyring1930}---to the mechanisms and kinetics of chemical reactions. Existence of multiple minima on the PES underpins concepts such as isomers, conformers and rotamers; and the curvature at minima supplies the means to interpret infra-red spectroscopy. In short, the Born-Oppenheimer approximation helps rationalize and explain many of the central concepts of modern chemistry.

Even so, the Born--Oppenheimer approximation breaks down in a wide range of chemically important scenarios: nonadiabatic effects play an important role in practically all photo-activated processes, in molecular electronics, and in electron-transfer reactions.\cite{Tully2012} Moreover from the theoretical viewpoint, the Born--Oppenheimer approach has a major drawback: it converts a problem with at worst two-body interactions into one in which the coordinates of all nuclei are coupled together through potential-energy surfaces.
Much of the ingenuity of the field of chemical quantum dynamics has been aimed at undoing or circumventing the complexities introduced by many-body potential energy surfaces.

For example, a number of approaches have been developed in a mixed quantum/classical framework, in which the nuclear degrees of freedom are treated through one or several classical trajectories evolving on potential energy surfaces, with corrections for nonadiabatic effects. 
The simplest case, Ehrenfest dynamics,\cite{Ehrenfest1927} arises from a mean-field treatment in which the nuclear  trajectory evolves on an averaged PES, weighted by excited-state populations.
In surface hopping,\cite{Tully1971,Tully1990}  classical trajectories are propagated under the forces of a single potential energy surface, but with stochastic hopping events between electronic states, allowing effective treatment of tunneling and nonadiabatic effects in averaged quantities.\cite{Parandekar2005,Tully2012}

Another class of methods involves propagation of quantum wavepackets in a basis that evolves through dynamics on some form of potential energy surface: in particular, such methods typically use a moving Gaussian basis to represent the nuclear wave packet. Since the introduction of this idea by Heller,\cite{Heller1975,Heller1981a,Heller1981} it has formed the basis of many modern developments in the nonadiabatic dynamics.
For example, in ab-initio multiple spawning,\cite{Martinez1996,Ben-Nun1998,Ben-Nun2000} classically moving Gaussian functions are evolved over potential energy surfaces, spawning new trajectory basis functions in highly nonadiabatic regions. Quantum dynamics in the moving evolving basis on different electronic states captures nonadiabatic effects, and includes interference effects between parts of the wavepacket that have split onto different surfaces. Numerous recent extensions and developments have made this a particulary powerful approach for nonadiabatic dynamics.\cite{Curchod2016,Mignolet2016,Curchod2020}

Variationally solving the time-dependent Schr\"odinger equation, using a basis of Gaussian nuclear wave packets lead to the variational multi-configurational Gaussian (vMCG) method of nonadiabatic dynamics.\cite{Richings2015}
In vMCG each of several coupled electronic states is described by a basis of multiple Gaussian functions.
The nonadiabatic coupling between electronic states determines the equations of motion for the expansion coefficients and some quantum mechanical parameters of the wavepackets,
while classical equations of motion are used for the position and momentum of thawed basis functions.

Considerable recent effort has been made to extend path integral dynamics methods to the nonadiabatic regime.
For example in the iso-RPMD extension to ring polymer molecular dynamics\cite{Tao2018}, the RPMD classical Hamiltonian is sampled over multiple potential energy surfaces using surface hopping, or evolves on an Ehrenfest averaged surface, producing correct thermodynamic properties while also approximately\cite{Lawrence2019} recovering nonadiabatic effects.
An alternative path-integral approach smoothly interpolates between the quantum instanton in the adiabatic limit and Wolynes's theory in the golden-rule limit of the nonadiabatic regime.\cite{Lawrence2020}

Finally, there are fully quantum methods based on wavefunctions that describe both electronic and vibrational degrees of freedom. Amongst these the time-dependent Hartree (TDH) method and particularly the multi-configurational generalization (MCTDH)  are particularly noteworthy.\cite{Meyer1990,Beck2000}
In MCTDH a superposition of electronic/nuclear product states is used, with nuclear functions described on grids.
As the grid has the same dimensionality as the underlying potential energy surface, such treatments are only amenable to small systems or those of reduced dimensionality. Recent advances include on-the-fly implementations which attempt to fit a global potential energy surface using local information.\cite{Richings2018}

Alternatively, MCTDH methods can be applied to model vibronic Hamiltonians, thereby bypassing the introduction of potential energy surfaces.\cite{Wang2013} A very recent work applied the MCTDH machinery with a second-quantization representation of the electrons avoiding the need for potential energy surfaces.\cite{Sasmal2020} The electronic single-particle basis in that work corresponds to mean-field orbitals on top of a simple diabatization strategy.

Not all approaches to the problem rely on the introduction of potential energy surfaces in their original formulation. For example the nuclear-electronic orbital (NEO) method takes a decidedly quantum chemical approach, shifting the normal Born--Oppenheimer separation to include light  nuclei on the same footing as electrons.\cite{Webb2002}
The NEO-DFT\cite{Chakraborty2008,Chakraborty2009} method in particular has been applied to studies of proton densities and geometries with some success.\cite{Yang2017,Brorsen2017}

In parallel with developments mainly in the chemistry community, electron-phonon interactions have been a part of the theoretical fabric of condensed-matter physics since the foundational work in that field.\cite{Bloch1929}
In molecular quantum chemistry the logical flow involves the introduction of adiabatic potential-energy surfaces followed by dynamics that captures nonadiabatic effects; but in the condensed-matter field the approach is different, starting from an assumption of harmonic oscillations in the lattice and linear couplings between nuclear positions and electronic degrees of freedom. 

While in the early days such a framework was used to motivate model Hamiltonians, such as those of Fr\"ohlich\cite{Frohlich1954} 
and Holstein,\cite{Holstein1959} much recent work has been performed in the ab initio context, working to include more subtle interactions between electrons and phonons, and to apply resulting methods to a host of more complex challenges in condensed matter physics.\cite{giustino2017} 
Much of the work in the condensed matter community uses DFT and Green's-function-based many-body corrections,
but very recently work on periodic\cite{McClain2017} and finite-temperature coupled-cluster theory\cite{White2018,White2020a} has been brought to bear to study electron-phonon couplings in a coupled-cluster framework.\cite{White2020}

Here we begin to explore how the typical electron-phonon framework can be derived in a quantum chemical context,  applied to nonadiabatic processes in molecules. The goal in doing so is to move away from methods that invoke potential-energy surfaces that couple many degrees of freedom together. Instead, the intention is to construct approximate Hamiltonians that capture key phenomena beyond the Born--Oppenheimer approximation, and use the established hierarchy of quantum chemistry methods to explore the dynamics of these Hamiltonians. There is clearly some overlap with existing attempts to model nonadiabatic effects through model vibronic Hamiltonians; but, as we will show, the approach set out here avoids the need to compute individual excited-state PESs, and avoids the need to diabatize them. 

In this paper we set out the basic formalism, derive a molecular electron-vibration model Hamiltonian, and show results from mean-field and linear response theory; we show through a simple model calculation how correlated theories on such model Hamiltonians will yield non-trivial nonadiabatic effects; and we demonstrate how vibronic model Hamiltonians can be simply parameterized based on a one-shot calculation at a single reference geometry.

\section{Theory}
The total molecular Hamiltonian is given by
\begin{equation}
    \op{H}_{\rm mol} = \op{H}_{\rm el}\ofX + \op{T}_{\rm nuc}
\end{equation}
where $\myvec{X}$ provides the nuclear coordinates in terms of displacements in normal modes computed at a reference geometry, $\myvec X=\myvec0$.

The domain of $\op{H}_{\rm mol}$ is a tensor product of Hilbert spaces for electronic and vibrational degrees of freedom; however, we plan to derive a theory in a standard quantum chemical framework, in which the one-particle electronic basis functions are atomic orbitals connected to atomic centres. For this reason we deal with basis functions of the form $\Psi_{IP}(\myvec x,\myvec X)=\Phi_I(\myvec x;\myvec X)\chi_P(\myvec X)$, where $\myvec x$ are the electronic coordinates. Here, $\Phi_I(\myvec x; \myvec X)$ is an electronic Slater determinant for a particular set of nuclear coordinates $\myvec X$, and $\chi_P\ofX$ is a product of vibrational wavefunctions for each mode $\modeone$,
\begin{equation}
    \chi_P\ofX =
    \prod_\modeone \chi^\modeone_{P_\modeone}\of{ X_\modeone}.
\end{equation}

The electronic determinant $\Phi_I\of{\myvec{x};\myvec{X}}$ is constructed from electronic orbitals $\phi_p(\myvec x; \myvec X)$ that are taken to be orthonormal for all values of $\myvec X$. Standard (geometry-dependent) creation and 
annihilation operators $a_p^{(\dagger)}(\myvec X)$ allow us to build the Slater determinant electronic basis:
\begin{equation}
\Phi_I(\myvec x; \myvec X) =
\bra{\mathbf x}
\prod_{p\in I} a_p^\dagger(\myvec X)\ket{~} \;. \end{equation}
We use a similar set of vibrational creation and annihilation operators for each vibrational mode, denoted $\hat b_n^{\modeone(\dagger)}$ for the $n$-th modal of the vibrational mode labelled $\modeone$, as suggested by Christiansen.\cite{Christiansen2004}$^,$\footnote{While we could have used standard bosonic operators here, our intention in making this specific choice is motivated by two considerations: first, we intend to explore models beyond the harmonic approximation and different options for the vibrational basis; and second this choice is particularly convenient in the sense that it provides a clear route for extending electronic quantum chemistry into the electron-vibrational domain.\cite{Mordovina2020}}

To set up a second-quantized Hamiltonian we have to be careful in tracking dependence on $\myvec X$. As an example, we can think about the electronic kinetic energy operator. The underlying operator $-\tfrac12\nabla^2$ has no dependence on $\myvec X$, but we introduce an $\myvec X$-dependent second-quantized operator (whose $\myvec X$-dependence disappears in the basis-set limit):
\begin{equation}
    -\tfrac12\nabla^2
    \rightarrow
    \sum_{pq} t_{pq}(\myvec X)a_p^\dagger(\myvec X)a_q(\myvec X)
    \label{eq:telec2nd}
\end{equation}
where
\begin{equation}
    t_{pq}(\myvec X) = \braket{\phi_p(\myvec X)|{-\tfrac12\nabla^2}|\phi_q(\myvec X)} \;.
\end{equation}
The $\myvec{X}$-dependence of the creation and annihilation operators stems solely from the orbitals in which they create or annihilate particles. Because these operators themselves will only be used to change one number string to another, the $\myvec X$-dependence has no consequence, and we will drop it.

To construct the total second-quantized Hamiltonian we consider a typical matrix element
\begin{equation}
    H_{IP,JQ} = \braket{\Phi_I(\myvec{X})\chi_P|\op{H}_{\rm mol}|\Phi_J(\myvec{X})\chi_Q}
\end{equation}
where the $\myvec X$-dependence of the \emph{electronic} basis functions is made explicit for clarity, although obviously the nuclear basis functions also have this dependence.

The integrations over the electronic and nuclear degrees of freedom can be performed in either order, and we choose to integrate over electronic coordinates first:
\begin{equation}
    H_{IP,JQ} = 
    \braket{\chi_P|
    \braket{\Phi_I(\myvec{X})|\op{H}_{\rm mol}|\Phi_J(\myvec{X})}
    |\chi_Q} \;.
\end{equation}
The $\myvec{X}$-dependence of the electronic states couples the integration over both sets of coordinates, preventing straightforward evaluation. In previous work by some of us, the issue is solved by numerical integration over the nuclear degrees of freedom.\cite{Sibaev2020} Here, in order to achieve a more scalable solution, the approach is to approximate the inner integral over electronic coordinates with a truncated Taylor expansion in $\myvec X$.

The inner integral is separated into its constituent terms:
\begin{equation}
\braket{\Phi_I(\myvec{X})|\op{H}_{\rm mol}|\Phi_J(\myvec{X})}\circ = \braket{\Phi_I(\myvec{X})|\op{H}_{\rm el}\of{\myvec{X}}|\Phi_J(\myvec{X})} + \braket{\Phi_I(\myvec{X})|\op{T}_{\rm nuc}|\Phi_J(\myvec{X})}\circ\; ,
\label{eq:inner_int}
\end{equation}
where the $\circ$ suffix serves to emphasize that the result is an operator that will still act on the vibrational ket wavefunction, even after the integration over electronic degrees of freedom.

The first term is recognisable as an element of the standard clamped-nucleus electronic Hamiltonian as a function of $\myvec X$.
To enable simple calculation of the total matrix element, we make a truncated expansion of this
term about $\myvec X=\myvec 0$:
\begin{equation}
    \braket{\Phi_I\ofX)|\op{H}_{\rm el}\ofX|\Phi_J\ofX} = 
    H^{\rm el}_{IJ}\ofX =
    H^{\rm el}_{IJ}\ofzero +
    \myvec{X}\cdot\left[\boldsymbol\nabla H^{\rm el}_{IJ}\right]\ofzero +
    \frac12 \myvec X^T\mymat K\myvec X +
    \mathcal{O}\of{\myvec{X}^2} \;.
\label{eq:Hel_IJ}\end{equation}
Here we chose to introduce a model harmonic potential with diagonal force-constant matrix $\mymat K$; this is not exact, so further quadratic terms contribute along with higher-order terms. To each order, the terms of the expansion are hermitian in both the electronic and vibrational space. 

Now the second term of Eq.~(\ref{eq:inner_int}) is investigated. It is a matrix element over
\begin{equation}
    \op{T}_{\rm nuc} = \sum_\modeone \frac{\op{P}_\modeone^2}{2\mu_\modeone} =
    -\hbar^2\sum_\modeone \frac{\nabla^2_\modeone}{2\mu_\modeone}
\end{equation}
where $\modeone$ labels a normal mode with reduced mass $\mu_\modeone$.
To simplify the notation we now abbreviate $\Phi_I(\myvec X)$ as $I$ and introduce
\begin{align}
\pi^\modeone_{IJ}\ofX &= \braket{I|\nabla_\modeone|J} \\
\Pi^\modeone_{IJ}\ofX &= \braket{\nabla_\modeone I|\nabla_\modeone J} .
\end{align}

The $\nabla^2_\modeone$ operator after integration over the electronic degrees of freedom becomes
\begin{align}
\braket{I|\nabla_\modeone^2|J} \circ &=
\braket{I|(\nabla_\modeone^2J)} +
2\pi^{\modeone}_{IJ}\ofX\nabla_\modeone  +
\braket{I|J}\nabla_\modeone^2  \\
&=
\left[\nabla_\modeone \pi^{\modeone}_{IJ}\ofX\right] - \Pi^{\modeone}_{IJ}\ofX+
2\pi^{\modeone}_{IJ}\ofX\nabla_\modeone +
\delta_{IJ} \nabla_\modeone^2 \label{eq:T}
\end{align}
Apart from the pre-factors, the last term is the nuclear kinetic energy operator acting on the nuclear basis functions. Note that $-\Pi^{\modeone}_{IJ}\ofX$ and $\delta_{IJ} \nabla_\modeone^2$ are individually hermitian in both electronic and vibrational space while $\left[\nabla_\modeone \pi^{\modeone}_{IJ}\ofX\right] + 2\pi^{\modeone}_{IJ}\ofX\cdot\nabla_\modeone$ is antihermitian in each subspace but hermitian overall.\cite{Sibaev2020} Any approximation should keep these symmetries in order to ensure real observables.

Up to this point, the above is in the same form as in Ref.~\citenum{Sibaev2020}. However, in that work the integration over vibrational space is performed numerically, leading to prohibitive computational cost for systems with a large number of vibrational modes. Here we replace the numerical integrations with analytic integrals of the Taylor expansion of the integrand, in analogy to the expansion of $H_{IJ}^{\rm el}\ofX$.

Taylor-expansion of the terms containing electronic integrals yields
\begin{equation}
    -\Pi^\modeone_{IJ}\of{\myvec{X}} =
     -\Pi^\modeone_{IJ}\of{\myvec{0}} -
      \myvec{X}\cdot\left[\boldsymbol\nabla\Pi^\modeone_{IJ}\right]\of{\myvec{0}} + \mathcal{O}\of{\myvec{X}^2}
\end{equation}
and
\begin{align}
    \left[\nabla_\modeone \pi^{\modeone}_{IJ}\ofX\right] &+ 2\pi^{\modeone}_{IJ}\ofX\cdot\nabla_\modeone \\
    \notag={}&
    \left[\nabla_\modeone \pi^{\modeone}_{IJ}\ofzero\right] +
    2\pi^{\modeone}_{IJ}\ofzero\nabla_\modeone +
    \nabla_\modeone \left[\myvec{X}\cdot\left[\boldsymbol{\nabla}\pi^{\modeone}_{IJ}\right]\ofzero\right] + 2\left[\myvec{X}\cdot\left[\boldsymbol{\nabla}\pi^{\modeone}_{IJ}\right]\ofzero\right]\nabla_\modeone + \mathcal{O}\of{\myvec{X}^2}
\end{align}
The first term of the above vanishes due to the derivative operator. Thus, the second term contains all zeroth order contributions. It is antihermitian in both electronic and vibrational space. From the third term, only the derivative along mode $\modeone$ survives:
\begin{equation}
    \nabla_\modeone \left[\myvec{X}\cdot\left[\boldsymbol{\nabla}\pi^{\modeone}_{IJ}\right]\ofzero\right] =
    \nabla_\modeone\sum_{\modetwo}  X_\modetwo \left[\nabla_\modetwo \pi^{\modeone}_{IJ}\right]\ofzero = \left[\nabla_\modeone \pi^{\modeone}_{IJ}\right]\ofzero
\end{equation}
Combined with the other first order term, the total first order contribution along mode $\modetwo = \modeone$ is
\begin{equation}\label{eq:first_order_pi}
\left[\nabla_\modeone\pi^{\modeone}_{IJ}\right]\ofzero(1 + 2 X_\modeone\nabla_\modeone)
\end{equation}
and the contribution along all modes $\modetwo \neq \modeone$ is
\begin{equation}
    2\left[\nabla_\modetwo\pi^{\modeone}_{IJ}\right]\ofzero X_\modetwo \nabla_\modeone.
\end{equation}
The latter expression is antihermitian along $\modeone$ and hermitian along all other modes and thus overall antihermitian in vibrational space (and also antihermitian in electronic space). Although it is not obvious, Eq.~\ref{eq:first_order_pi} is antihermitian in vibrational space as well (see supplementary material).

As a conclusion, all above expressions keep the aforementioned symmetries separately for the zeroth-order and first-order terms. By inspection of Eq.~\ref{eq:T}, one may be tempted to Taylor-expand the expression \emph{after} application of the derivative operator in the first term. However, this would break the symmetries for the individual orders of the expansion.

As a result of the Taylor expansion, the integrations over electronic and vibrational degrees of freedom can now be performed separately. All integrals in the vibrational space are straightforward. The integrals in the electronic space result from the application of $\op{T}_{\rm nuc}$; application of the Slater--Condon rules allows $\pi^{\modeone}_{IJ}$ and $\Pi^\modeone_{IJ}$ to be obtained from the corresponding orbital integrals,
\begin{align}
   \pi^\modeone_{pq}\ofX  &=
   \braket{\phi_p\ofX|\nabla_\modeone|\phi_q\ofX} \\
   \Pi^\modeone_{pq}\ofX  &=
   \braket{\nabla_\modeone\phi_p\ofX|\nabla_\modeone\phi_q\ofX}
   \;.
\end{align}

So far, the orbital basis $\left\{\phi_p\of{\myvec{X}}\right\}$ has not been specified other than being orthonormal for all $\myvec{X}$. For molecular systems the basis functions are usually atom-centered, and so depend on   $\myvec{X}$. The electronic orbital basis is expanded in such atomic basis functions $\left\{\mu\of{\myvec{X}}\right\}$. The orbital basis is defined by an \myvec{X}-dependent transformation matrix \mymat{T}. Its \myvec{X}-dependence is vital in order to ensure orthonormality for all $\myvec{X}$. Interpreting $\left\{\ket{\mu\ofX}\right\}$ as a row vector of all atom-centred basis functions, the transformation to the orbital basis is given by
\begin{equation}
    \left\{\ket{\phi_p\of{\myvec{X}}}\right\} = \left\{\ket{\mu\of{\myvec{X}}}\right\}\mymat{T}\of{\myvec{X}}.
\end{equation}

The notation is now further simplified by implying that a dropped position-dependence means evaluation at $\myvec{0}$. The $n$-th derivative with respect to the coordinates of the $\modeone$-th vibrational mode of any object $A$, evaluated at $\myvec{0}$, is  denoted $A^{(n)}$. Furthermore, we write
$\mymat{S}^{(m, n)}=\braket{\mu^{(m)}|\mu^{(n)}}$ so that for example 
$\mymat S=\mymat{S}^{(0, 0)}$.

The integral matrices over electronic coordinates needed for the zeroth-order terms in the expansions become
\begin{align}
    \boldsymbol \pi_\modeone &=
    \mymat{T}^\dagger\mymat S^{(0, 1)}\mymat{T} +
\mymat{T}^\dagger\mymat S \mymat{T}^{(1)}
\\
    \mymat \Pi_\modeone &=
    \mymat{T}^{(1)\dagger}\mymat{S}^{(0, 1)}\mymat{T} +
\mymat{T}^{(1)\dagger}\mymat S\mymat{T}^{(1)} +
\mymat{T}^\dagger\mymat{S}^{(1, 1)}\mymat{T} +
\mymat{T}^\dagger\mymat S^{(1, 0)}\mymat{T}^{(1)} \; .
\end{align}
Analytic derivatives of the above with respect to mode $\modeone$, and also the other modes, (leading to mode-mode coupling) are straightforward if the corresponding derivatives of the transformation matrix are available. The first-order expressions along $\modeone$ are shown in the supplementary material.

\subsection{Electronic basis}
If the Taylor expansion includes all orders, and in the limit of a complete basis, the full molecular Hamiltonian is recovered and results become invariant to the choice of the electronic basis. The same holds for the underlying single-particle basis, so that results are invariant to the choice of $\mymat{T}\ofX$. 
However, when the Taylor expansion is truncated, different choices of $\mymat{T}\ofX$ do lead to different results, and it becomes important to consider how different basis sets perform in the context of these approximations. Before setting out the approach we take here, it is worth noting
that this issue has also been considered in perturbation theory under the heading of orbital connection.\cite{Olsen1995, Ruud1995}

One obvious and intuitively reasonable choice for $\mymat T(\mymat X)$ is the coefficient matrix $\mymat{C\ofX}$ of the optimized mean-field orbitals at each \myvec{X}, which is also used in Ref.~\citenum{Sibaev2020}. This choice amounts to a kind of one-particle adiabatic basis: the coupling Hamiltonian has an electronic part that is diagonal at each value of \myvec{X}. No coupling is induced \emph{between} different Slater determinants constructed from mean-field orbitals through the one-particle (Fock matrix) approximation to $\hat{H}_{\rm el}\ofX$. It will be shown further below, that this choice in fact has dramatic consequences on low-order approximations turning the seeming advantage of a diagonal electronic Hamiltonian into a severe disadvantage. Clearly, in situations where the above approximations are valid, the coupling via $\hat{H}_{\rm el}\ofX$ will be small for this basis choice. Another drawback of the mean-field orbital basis in the context of this work is the computational expense for the calculation of derivatives of the transformation matrix.

The diabatic basis suggested by Troisi and Orlandi\cite{Troisi2003} does not suffer from these drawbacks. Here, only derivatives of $\mymat{S}\ofX$ (and of its inverse, in case of higher derivatives) are required in the calculation of derivatives of the transformation matrix. Furthermore, it has the appealing advantage that by construction $\boldsymbol{\pi}\ofzero$ vanishes. It should be emphasized, though, that the one-particle diabatization that leads to this choice does not amount to diabatization of the many-particle states; indeed such a diabatization is not generally possible.\cite{Mead1982}

The aforementioned basis sets focus on the change of molecular orbitals upon displacement. 
In quantum chemistry we typically use nonorthogonal atom-centred basis functions, and so a large part of the electron-nucleus coupling simply arises from the changing  metric, which in turn leads to Pulay forces, which are known to be far from negligible. The bases listed above conflate the issue of a changing metric with changes induced by actual physical electron-nuclear coupling effects. 

To avoid this we also investigated the simplest choice for $\mymat{T}\ofX$ that resolves the issue of the changing metric whilst leaving the basis functions as close as possible to the original atomic orbitals. That is, we use the symmetric orthogonalized basis, with $\mymat{T}\ofX = \mymat{S}^{-\frac{1}{2}}\of{\myvec{X}}.$\footnote{In order to achieve a more seamless integration into existing electronic structure code, we perform our calculations in the AO basis at reference position which leads to an additional right factor of $\mymat{S}^{+\frac{1}{2}}\ofzero$. This choice does not affect any results.} 
This choice correctly deals with the changing metric but does not induce any further rotation of the orbital basis upon displacement. Effectively, this basis is identical to a frozen orbitals basis, which has been shown to represent an excellent choice for a quasi-diabatic basis in case one wants to avoid the explicit calculation of derivative couplings.\cite{Pacher1991} In stark contrast to using mean-field orbitals, the main vibronic coupling effect is captured through $\hat{H}_{\rm el}\ofX$, because all changes to the electronic state caused by moving nuclei (other than those resulting from a changed metric) have to be made through explicit orbital rotations. Analytic expressions for the derivatives of $\mymat{S}^{-\frac{1}{2}}\of{\myvec{X}}$ are given in the supplementary material.

At the level of approximation we employ in this work, we found that none of the other bases mentioned above are competitive with the symmetrically orthogonalized basis; this is demonstrated below through comparisons of the three discussed choices for $\mymat{T}\ofX$. It also accords with the simple intuitive picture afforded by atomic-orbital basis sets: properties of diabatic states should vary smoothly with geometry, and this seems to be a clear attribute of atomic orbitals that follow nuclear positions in a straightforward way. The choice of $\mymat{T}\ofX= \mymat{S}^{-\frac{1}{2}}\of{\myvec{X}}$ remains as close as possible to this intuitively simple picture, while ensuring that the orbitals are orthogonal at all geometries.

\subsection{Approximations}
We apply the following set of approximations:
\begin{enumerate}
    \item All second-order terms are replaced by a harmonic potential with a fixed force constant in each mode (see Eq.~\ref{eq:Hel_IJ}).
    \item Only the zeroth-order terms of Taylor expansions of $\boldsymbol{\pi}\ofX$ and $\mymat{\Pi}\ofX$ are included.\label{it:pi}
    \item The gradient of $\hat{H}_{\rm el}\ofX$ is replaced by the gradient of a mean-field approximation to avoid three-body terms (see below)\label{it:Hel}.
\end{enumerate}
Each approximation can be separately improved on if necessary. Preliminary calculations showed that due to the scaling with the inverse mass, contributions from coupling via $\op{T}_{\rm nuc}$ are small and the first order terms neglected in Item~\ref{it:pi} are significantly smaller than the zeroth order terms. Item~\ref{it:Hel} embodies some peculiarities which are discussed in the following.

The first-order term of the electronic Hamiltonian, $\myvec{X}\cdot\left[\boldsymbol\nabla H^{\rm el}_{IJ}\right]\ofzero$, contains three-body terms involving the two-particle electronic integrals and the one-particle displacement operator. These are prohibitive for an efficient solution of the Schr\"odinger equation, so we replace the coupling with a mean-field approximation. While this is routinely done in the condensed-matter literature, we here set out a derivation that illustrates the nature of the approximation.  

Adding and subtracting the expectation value for the mean field ground state at any $\myvec{X}$ yields
\begin{eqnarray}
    \nabla_\modeone\hat{H}_{\rm el}\ofX &=&
    \langle\Phi_\mathrm{mf}|\nabla_\modeone\hat{H}_{\rm el}\ofX|\Phi_\mathrm{mf}\rangle  +
    \nabla_\modeone\hat{H}_{\rm el}\ofX -
    \langle\Phi_\mathrm{mf}|\nabla_\modeone\hat{H}_{\rm el}\ofX|\Phi_\mathrm{mf}\rangle \\
    &=&
    \nabla_\modeone E_{\rm mf}^{\rm el}\ofX +
    \nabla_\modeone\hat{H}_{\rm el}\ofX -
        \langle\nabla_\modeone\hat{H}_{\rm el}\ofX\rangle\nonumber,
\end{eqnarray}
where $E^{\rm el}_{\rm mf}$ is the mean-field ground state electronic energy, and where a notational simplification is introduced for the expectation value with the mean-field ground state. The second equality arises because the mean-field state is variationally optimized at each geometry.
The electronic Hamiltonian can be split up into the Fock operator $\hat F$ and a fluctuation operator $\hat{V}$, commonly used in perturbation theory. Thus,
\begin{equation}
    \nabla_\modeone\hat{H}_{\rm el} = \nabla_\modeone E^{\rm el}_{\rm mf} + 
    \nabla_\modeone\hat{F} - \langle\nabla_\modeone\hat{F}\rangle + 
    \nabla_\modeone\hat{V} - \langle\nabla_\modeone\hat{V}\rangle,
\end{equation}
where the $\myvec{X}$-dependence is omitted for further brevity.
We then neglect the last two terms (which contain the 3-body contribution) to yield
\begin{equation}
    \nabla_\modeone\hat{H}_{\rm el} \approx \nabla_\modeone E^{\rm el}_{\rm mf} + 
    \nabla_\modeone\hat{F} - \langle\nabla_\modeone\hat{F}\rangle \,
\end{equation}
an approximation that should be valid for states whose densities are not too different from the mean-field ground-state density.

The Hamiltonian gradient in the electron-vibration coupling is then given by the one-electron operator
\begin{equation}
    \nabla_\modeone\hat{H}_{\rm el} = \sum_{pq} [\mymat{A_\modeone}]_{pq}\hat a_p^\dagger\hat a_q + c_\modeone
\end{equation}
where
$\mymat{A}_\modeone = [\nabla_\modeone\mymat{F}]\ofzero$ and $c_\modeone = \nabla_\modeone E_{\rm mf}^{\rm el}\ofzero -
{\rm tr}(\mymat{D}_{\rm el}^{\dagger}\mymat{A}_\modeone)$. Here the constant term $c_\modeone$ reflects the fact that the gradient of the mean-field energy at the reference geometry need not match the matrix element of the derivative of the fock operator.

\subsection{Analysis of the linear coupling matrix \mymat{A}}
In this section, we only consider a single mode and thus omit the mode index $\modeone$. Within a unitary transformation, the Fock matrix, $\mymat{F}_{\rm el}\!\left[\mymat{D}_{\rm mf}^{\rm el}\ofX\right]$, does not depend on the choice of the position-dependence of the electronic basis, i.e.\@ on $\mymat{T\ofX}$. Still, the first-order coupling matrix \mymat{A}, which is its first derivative, \emph{does} depend on $\mymat{T\ofX}$. This can be seen most easily when comparing the mean-field orbital basis ($\mymat{T}\ofX = \mymat{C}\ofX$) to any other choice. In the mean-field orbital basis, the Fock matrix is the diagonal matrix of orbital eigenvalues for all $\myvec{X}$, thus the off-diagonal elements and all their derivatives are zero. A linear Taylor expansion is then identical to a linear expansion of the orbital energies. For any other choice of $\mymat{T}\ofX$, the Fock matrix is \emph{not} diagonal for all \myvec{X}. Still, approximate orbital energies can be obtained by diagonalizing a Taylor expansion of the Fock matrix in \emph{that} basis. However, the eigenvalues of a matrix do not depend linearly on the values in the off-diagonal elements. Thus, the orbital energies will in this case \emph{not} depend linearly on \myvec{X}. Clearly, in an infinite-order Taylor expansion, the same eigenvalues are recovered for any choice of $\mymat{T}\ofX$. Thus, the choice of $\mymat{T}\ofX$ leads to different convergence behaviour of a Taylor-expansion of the Fock matrix.

It is now evident that in the mean-field orbital basis no coupling between different orbitals is possible via $\mymat{A}$. In contrast, any other choice of $\mymat{T}\ofX$ will usually lead to couplings (via $\mymat{A}$) between almost all orbital pairs for which it is not avoided by symmetries of the system under investigation. As a consequence, couplings between different states are ubiquitous. Note that this is crucially different from approaches in which the electronic Hamiltonian and its derivatives are evaluated separately from the vibrational degrees of freedom. In such computations, each coupling between a pair of states must be considered explicitly.

The ubiquitous coupling between states includes couplings to high-energy states which are poorly described by the truncated model Hamiltonian. Preliminary calculations revealed instabilities resulting from such couplings, which lead to convergence problems and unphysical results.

Often only a limited number of excited states, and most often only singly excited states, are of relevance for nonadiabatic calculations. Thus, a large fraction of the information contained in \mymat{A} is never required, including those parts that are the root cause of the problems in practical calculations. We have therefore developed a strategy to project out all problematic couplings, which we here describe taking $\mymat{A}$ to be in the molecular orbital basis at the reference position.

The diagonal elements do not couple different orbitals to each other, but do play an important role in determining displacements in excited-state minima, and are fully retained.

The occupied-virtual block of the matrix can be understood as a vector in the space of single-particle excitations.  
Thus, the occupied-virtual block of $\mymat{A}$ can be projected onto the subspace of the single-particle excitation vectors corresponding to the states of interest:
\begin{equation}
    \widetilde{\mymat{A}}_{\rm ov} =  \mymat{P}\mymat{A}_{\rm ov},
\end{equation}
where $\mymat{A}_{\rm ov}$ is the occupied-virtual block of $\mymat{A}$, flattened out as a vector in the space of single-particle transitions, and
the projector \mymat{P} is
\begin{equation}
    \mymat{P} = \mymat{X}_{\rm lr}\mymat{X}_{\rm lr}^{\rm T} - \mymat{Y}_{\rm lr}\mymat{Y}_{\rm lr}^{\rm T},
\end{equation}
with the matrix of relevant excitation vectors $\mymat{X}_{\rm lr}$ and de-excitation vectors $\mymat{Y}_{\rm lr}$ which are obtained from time-dependent linear-response Hartree--Fock (or Kohn--Sham) calculations. (If the Tamm-Dancoff approximation is used, the $\mymat{Y}_{\rm lr}$ term is omitted.) 

The above projection makes use of single excitations only. Thus, in case states which are not dominated by single-particle transitions are of high relevance, this procedure would need to be adapted. We want to point out, though, that the strategy employed here does not necessarily lead to a bad description of multiply excited states in the approximate molecular Hamiltonian.

Often, all relevant excitations lie within the valence space. At the same time, excitations from core orbitals or into high virtual orbitals correspond to high-energy excitations and may thus be a main cause of the observed problems. Thus, all couplings outside the valence region are omitted.

The off-diagonal elements in the occupied-occupied and virtual-virtual blocks of \mymat{A} are the leading-order contribution to couplings between excited states. Unlike the occupied-virtual block, these blocks are not defined in the same space as the states they couple at the lowest order. In contrast, here a single matrix element is relevant for couplings between a large number of pairs of states. Thus, a strict separation of couplings between states of interest and couplings between states of less or no relevance is not possible within the coupling matrix in principle. Still, at least in the virtual-virtual blocks, matrix elements with (to lowest order) no relevance for any of the states of interest can be discarded, i.e.\@ all elements $A_{ab}$ for which
\begin{equation}
    |X^{n}_{ia} X^{m}_{jb}| < {\rm thr}, \forall \ i, j, n, m
\end{equation}
where currently a threshold of $10^{-6}$ a.u.\@ is chosen. Hereby, $X^{n}_{ia}$ is the element in the excitation vector for state $n$ representing the excitation of a particle from orbital $i$ into orbital $a$.
Orbitals in the occupied-occupied block are highly important and couplings between them may be important for orbital relaxation; they are therefore retained.

\subsection{Mean-field theory for the coupled electron-vibration Hamiltonian}
We begin our exploration of quantum chemical methods for the coupled electron-vibrational problem at the simplest, mean-field level.
Up to now, only the electronic terms have been quantized. In the following, also the vibrational terms will be used in second quantized form, so that
\begin{equation}
    X_\modeone \rightarrow \sum_{mn} X^{\modeone}_{mn}b^\dagger_m b_n,
\end{equation}
where the double sum is taken over modals $m, n$, and
\begin{equation}
    X^{\modeone}_{mn} = \braket{m|\hat{X}_\sigma|n}\;.
\end{equation}
The matrix representation $\boldsymbol\nabla_\modeone$ of the gradient operator in the vibrational basis is obtained analogously.

The interaction term between the electronic and vibrational subsystems is
\begin{equation}
    \mymat{H}_{\rm int} =
    (\mymat{A}_\modeone + c_\modeone\mymat{1})\otimes\mymat{X}_\modeone - \frac{\hbar^2}{2\mu_\modeone}\left( -\mymat{\Pi}_\modeone\otimes\mymat{1} + 2 \boldsymbol{\pi}_\modeone\otimes\boldsymbol\nabla_\modeone\right),
    \label{eq:H_int}
\end{equation}
where summation over $\modeone$ is implied.

Each of the above terms represent a tensor product of one-particle integrals in electronic space and one-particle integrals in vibrational space. The mean-field interaction energy expression is thus easily obtained by tracing with the corresponding density matrices ($\mymat{D}_\modeone$ is the density matrix of mode $\modeone$):
\begin{align}
    E_{\rm int} &= ({\rm tr}(\mymat{D}_{\rm el}^\dagger\mymat{A}_\modeone) + c_\modeone){\rm tr}(\mymat{D_\modeone^\dagger}\mymat{X}_\modeone)\\
    &- \frac{\hbar^2}{2\mu_\modeone}\left[ -{\rm tr}(\mymat{D}_{\rm el}^\dagger\mymat{\Pi}_\modeone) + 2 {\rm tr}(\mymat{D}_{\rm el}^\dagger\boldsymbol{\pi_\modeone}){\rm tr}(\mymat{D}_\modeone^\dagger\boldsymbol{\nabla}_\modeone) \right]\nonumber,
\end{align}
where it has been used that ${\rm tr}(\mymat{D}_\modeone) = 1$.

The Fock (or Kohn--Sham) matrix contributions due to the interaction are the derivatives of the above with respect to the corresponding density matrices,
\begin{equation}
    \mymat{F}^{\rm int}_{\rm el} = \mymat{A}_\modeone{\rm tr}(\mymat{D_\modeone^\dagger}\mymat{X}_\modeone) - \frac{\hbar^2}{2\mu_\modeone}\left[ -\mymat{\Pi}_\modeone + 2 \boldsymbol{\pi}_\modeone{\rm tr}(\mymat{D}_\modeone^\dagger{\boldsymbol\nabla}_\modeone) \right],
\end{equation}
and
\begin{equation}
    \mymat{F}^{\rm int}_\modeone = ({\rm tr}(\mymat{D}_{\rm el}^\dagger\mymat{A}_\modeone) + c_\sigma)\mymat{X}_\modeone - \frac{\hbar^2}{\mu_\modeone}
    {\rm tr}(\mymat{D}_{\rm el}^\dagger\boldsymbol{\pi}_\modeone){\boldsymbol\nabla}_\modeone.
\end{equation}

The energy contribution due to the term containing $\mymat{\Pi}$ does not depend on the vibrational density, so it can be included into the electronic core Hamiltonian. It resembles part of the diagonal Born--Oppenheimer correction (DBOC).

The Fock matrix for each vibrational mode is just the sum of the harmonic oscillator Hamiltonian and the above interaction term,
\begin{equation}
    \mymat{F}_\modeone = \mymat{H}^{\rm HO}_\modeone + \mymat{F}_\modeone^{\rm int}.
\end{equation}

Given the expressions for the Fock matrices, applying a coupled self-consistent field procedure is not much different from spin-unrestricted electronic mean-field calculations, where separate Fock and density matrices are used for each spin. Here, for each vibrational mode an additional Fock and density matrix appears. Convergence acceleration schemes typically used in electronic self-consistent field algorithms like Pulay's direct inversion of the iterative subspace\cite{Pulay1982} (DIIS) can be straightforwardly applied. Although more specialized schemes could be developed, we found that common DIIS variants\cite{Hu2010} work sufficiently well in the context of the coupled calculations.

\subsection{Coupled time-dependent linear response theory}
In electronic structure theory, the time-dependent linear response framework is the most widely used method for calculating excited states.\cite{Dreuw2005, Furche2005} Starting from a coupled mean-field solution, that framework can be straightforwardly used in this context, especially in conjunction with iterative solvers like the Davidson solver.\cite{Davidson1975} Apart from a contribution from the orbital energies, the response matrix consists of the occupied-virtual occupied-virtual (plus the occupied-virtual virtual-occupied) block of the second derivative of the energy with respect to the density matrix, which is the derivative of the Fock matrix with respect to the density. In iterative schemes, the response matrix is not explicitly constructed, but the product of the response matrix with a guess transition density is calculated directly. This is (apart from the orbital energy contribution) essentially identical to a multiplication of the density derivative of the Fock matrix with the density. The interaction Fock matrices for both the electronic and vibrational subspace only have up to a linear dependence on the density (of the respectively other subspace). These terms are identical in the Fock matrix and the product of the response matrix with a guess transition density where each density matrix is just replaced by the corresponding transition density matrix. The constant terms only contribute via the change of the orbital energies.

In our framework it is thus straightforwardly possible to incorporate the coupling to vibrations into existing electronic structure programs. This has been shown for self-consistent field as well as for time-dependent linear response calculations and can be expected for established correlated electronic structure methods, too.

\section{Results and Discussion}
\subsection{Computational Details}
All calculations in this section have been performed with a development version of the Entos Qcore package,\cite{qcore}$^,$\footnote{Entos Inc., http://entos.ai, accessed Oct 6th, 2020}%
 except for the propagation calculations, which have been performed in \textsc{Mathematica}.\cite{Mathematica} The PBE0 functional\cite{Adamo1999} in the Def2-TZVP electronic basis set\cite{Weigend2005} is employed throughout. Density fitting has been used for Coulomb and exchange contributions\cite{Fruechtl1997} with the fitting basis corresponding to the atomic orbital basis set.\cite{Weigend2008}

\subsection{Potential energy surfaces}
We have presented an approximation to the molecular Hamiltonian, and we would like to establish its accuracy. While the theoretical direction we are taking eliminates the need for potential energy surfaces, they are nevertheless an important means by which to test the accuracy of the model Hamiltonian; for this reason 
alone, we now invoke the BO approximation, removing the nuclear kinetic energy operator, other nuclear-mass-dependent terms, and interpreting  $\mymat X$  as a classical variable. In this approximation the interaction Hamiltonian (for mode $\modeone$) becomes
\begin{equation}
    \mymat{H}_{\rm int}^{\rm BO}(X_\modeone) =
    (\mymat{A}_\modeone + c_\modeone\mymat{1})X_\modeone \;.
    \label{eq:H_int_BO}
\end{equation}
In practice, the ground-state potential energy surface is obtained from a standard mean-field electronic structure calculation
with the core Hamiltonian modified by the addition of $\mymat{H}_{\rm int}^{\rm BO}(X_\modeone)$, plus the harmonic potential $kX_\modeone^2/2$.
Excited states are obtained from corresponding electronic linear response calculations.

The pyrazine molecule has become a guinea pig for studies of nonadiabatic effects and has been subject to numerous computational studies.\cite{Seidner1992, Woywod1994, Raab1999, Sala2015} In contrast to diatoms or other small molecules, here the harmonic model potential can be expected to represent a good approximation. In pyrazine, the lowest singlet states of B$_{3\mathrm{u}}$($\mathrm{n}\pi^*$) and B$_{2\mathrm{u}}$($\pi\pi^*$) symmetry feature a conical intersection which leads to significant broadening of the B$_{2\mathrm{u}}$ peak in the absorption spectrum. To first order, these states are coupled only by the $\nu_{10a}$ mode which is the only mode of B$_{1\mathrm{g}}$ symmetry. The short-term dynamics of the system after excitation to the B$_{2\mathrm{u}}$ state is mostly governed by the coupling to the B$_{3\mathrm{u}}$ state due to the $\nu_{10a}$ mode and by the totally symmetric tuning modes.

\begin{figure}
    \centering
    \begin{subfigure}{.3\textwidth}
    \includegraphics[width=\textwidth]{"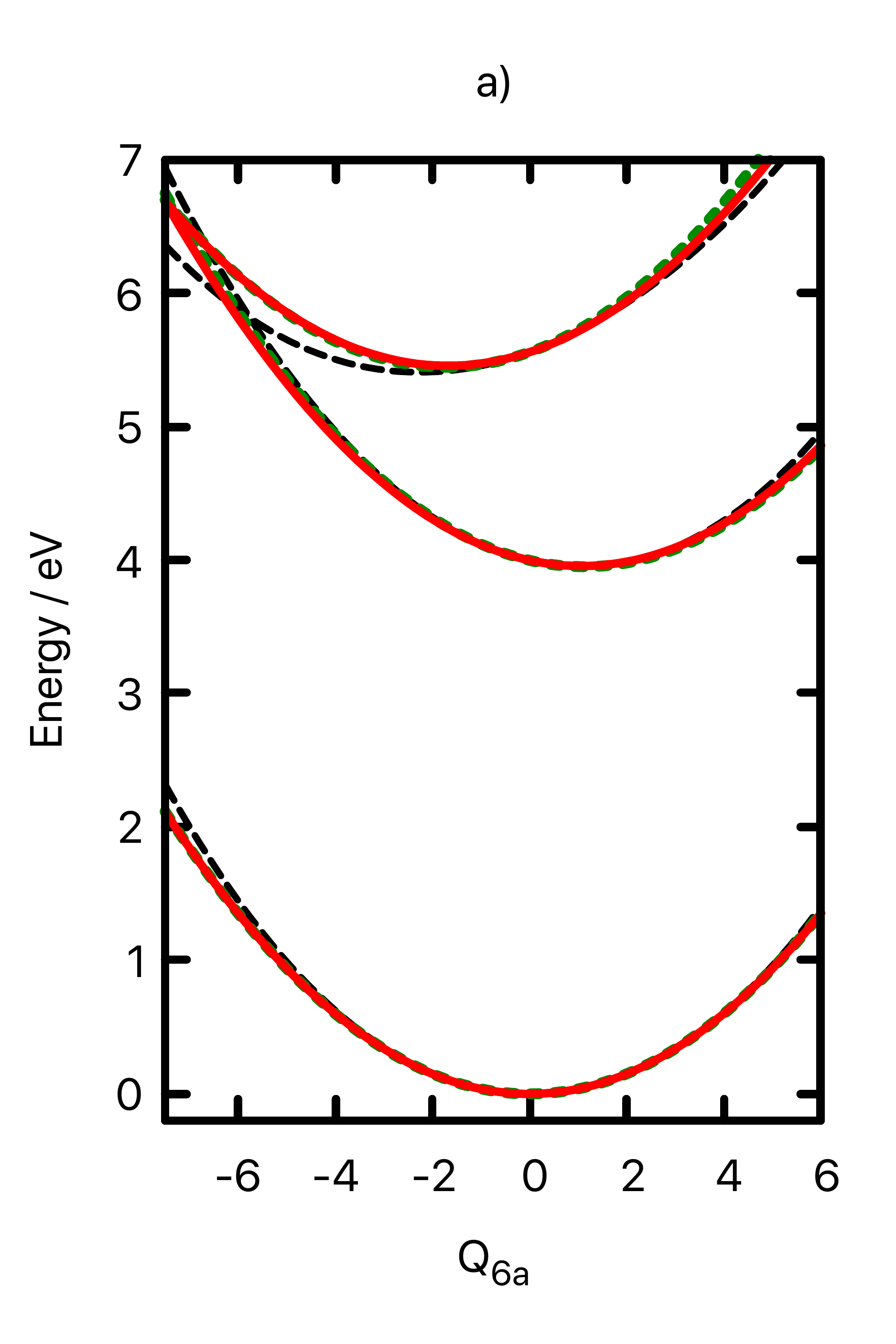"}
    \label{fig:pyrazine_PES_0}
    \end{subfigure}
    \begin{subfigure}{.3\textwidth}
    \includegraphics[width=\textwidth]{"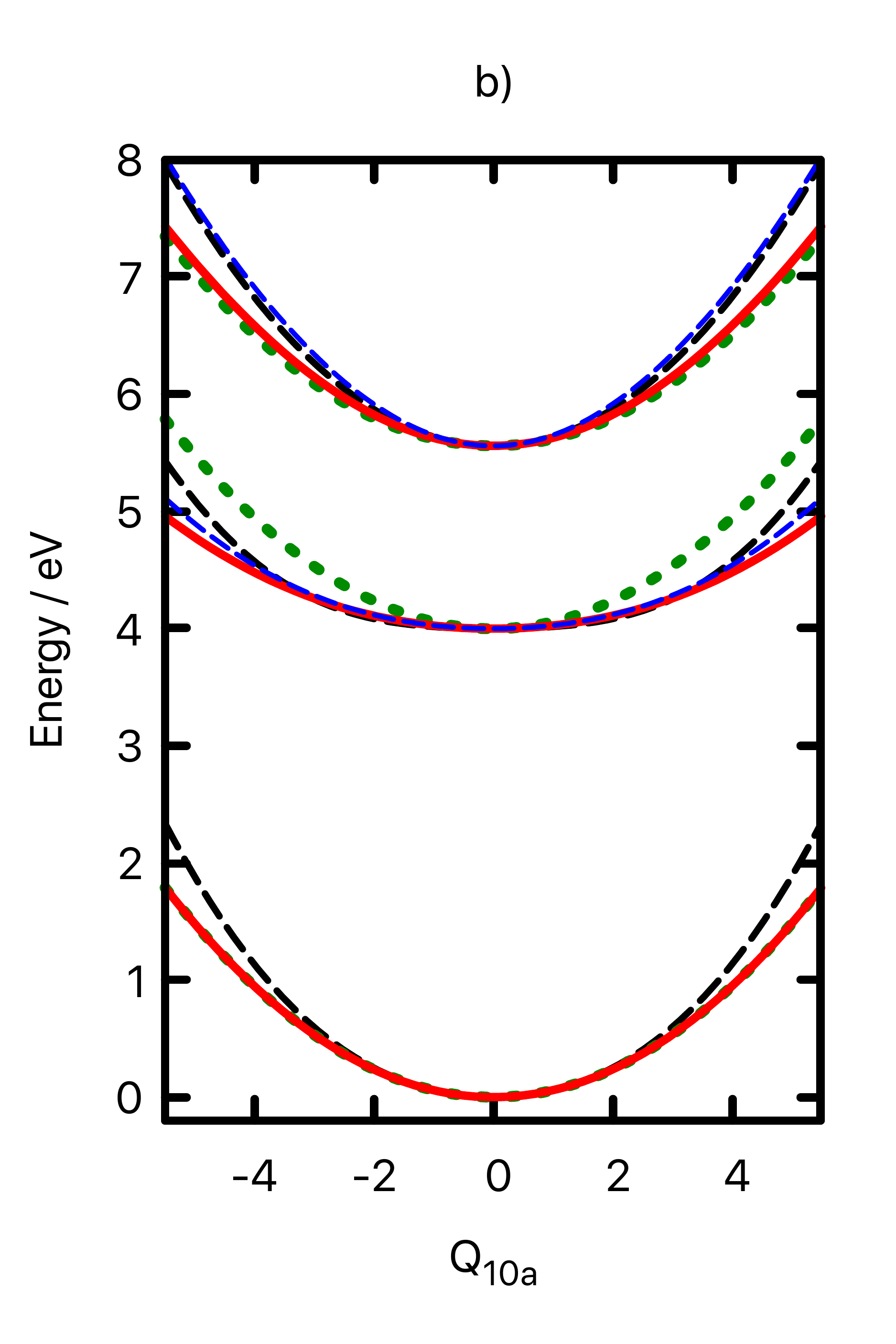"}
    \label{fig:pyrazine_PES_1}
    \end{subfigure}
    \begin{subfigure}{.3\textwidth}
    \includegraphics[width=\textwidth]{"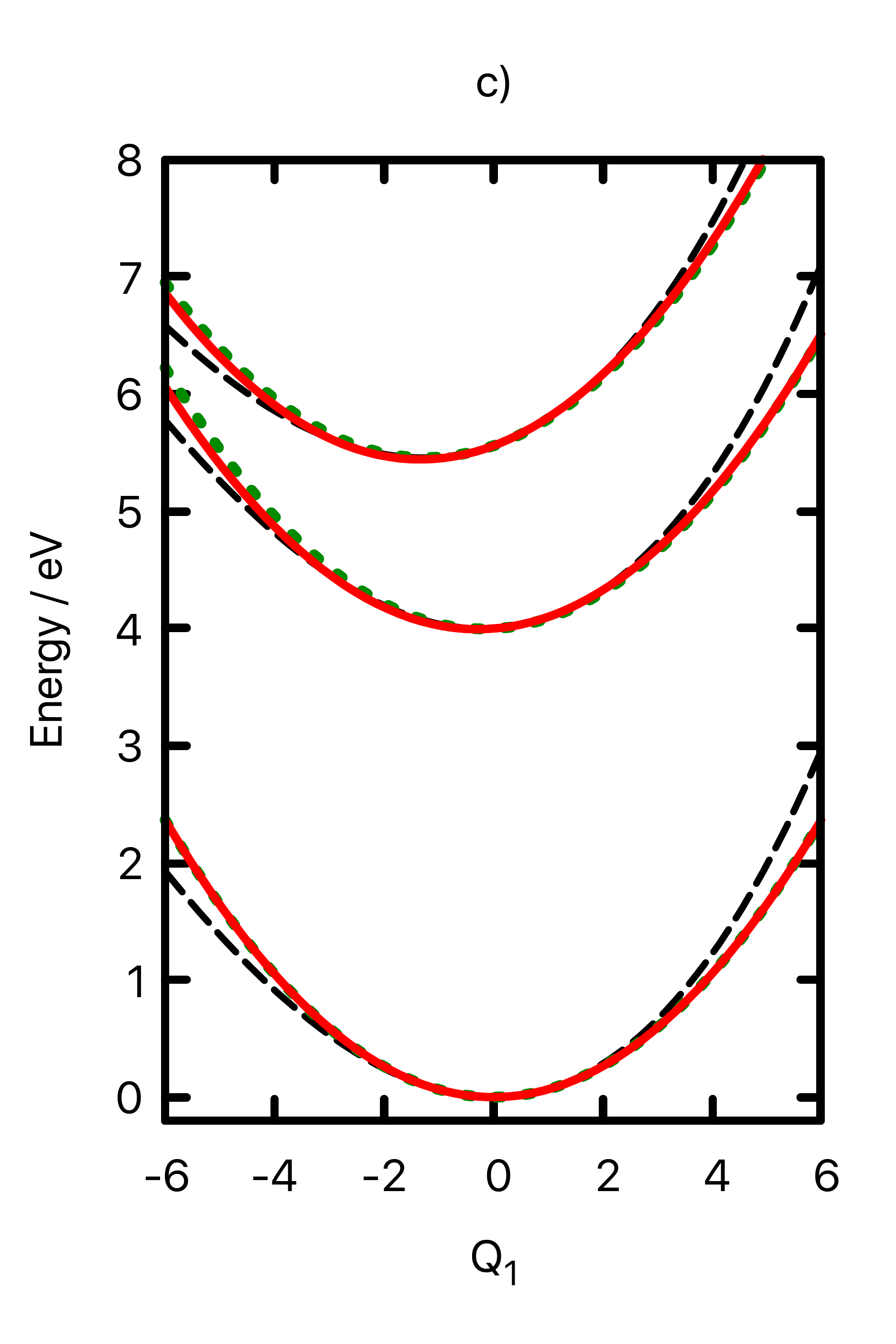"}
    \label{fig:pyrazine_PES_2}
    \end{subfigure}
    \begin{subfigure}{.3\textwidth}
    \includegraphics[width=\textwidth]{"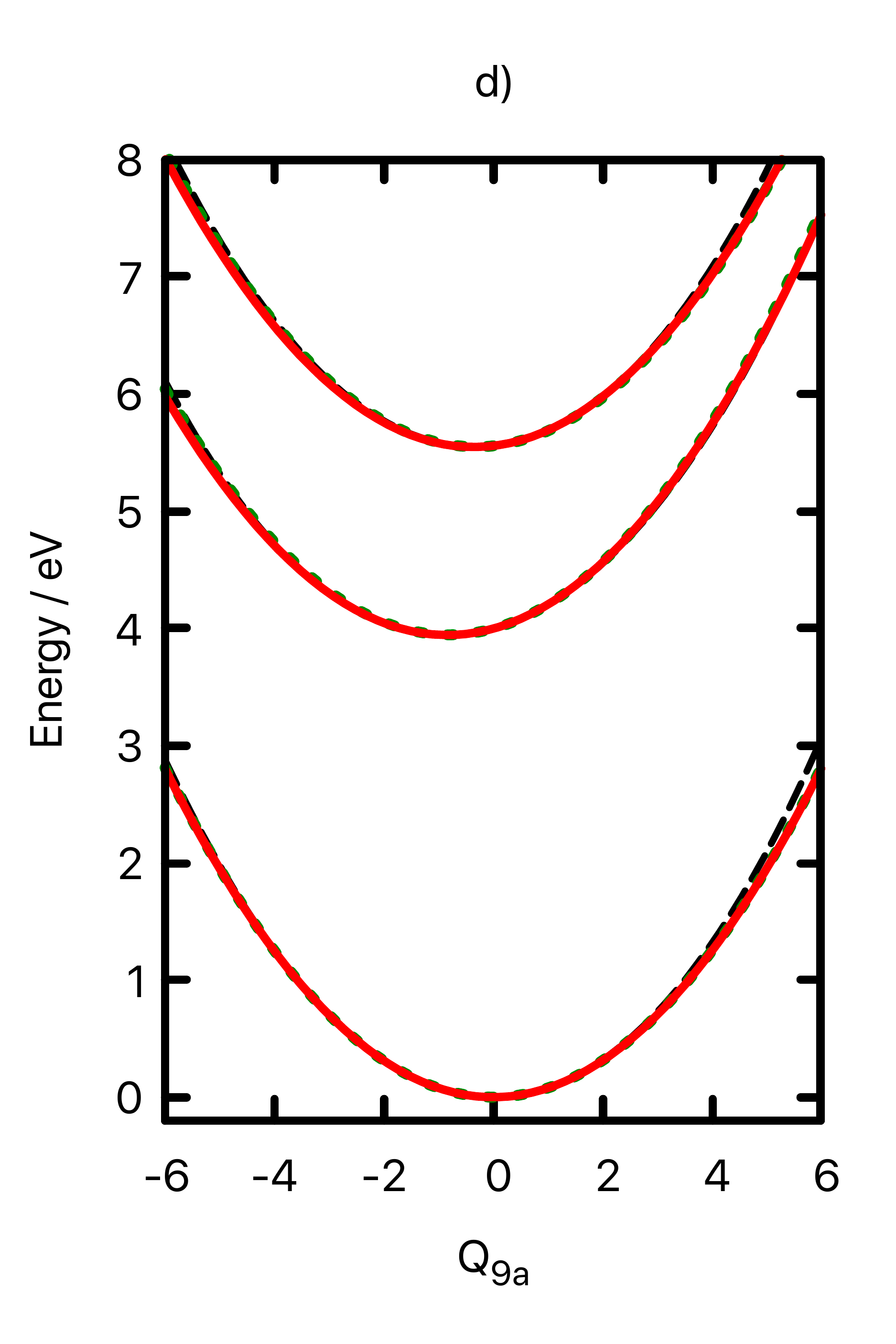"}
    \label{fig:pyrazine_PES_3}
    \end{subfigure}
    \begin{subfigure}{.3\textwidth}
    \includegraphics[width=\textwidth]{"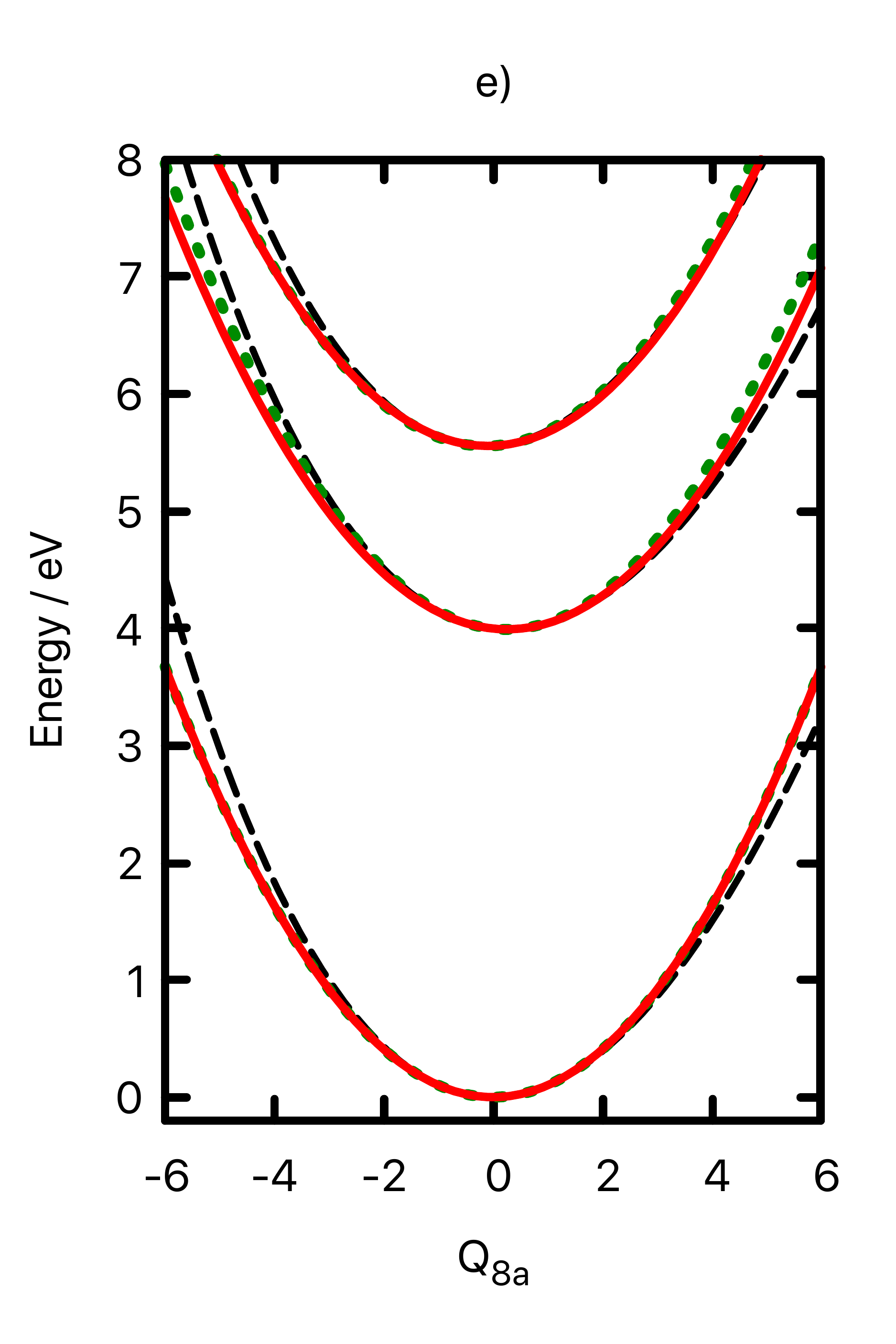"}
    \label{fig:pyrazine_PES_4}
    \end{subfigure}
    \begin{subfigure}{.3\textwidth}
    \includegraphics[width=\textwidth]{"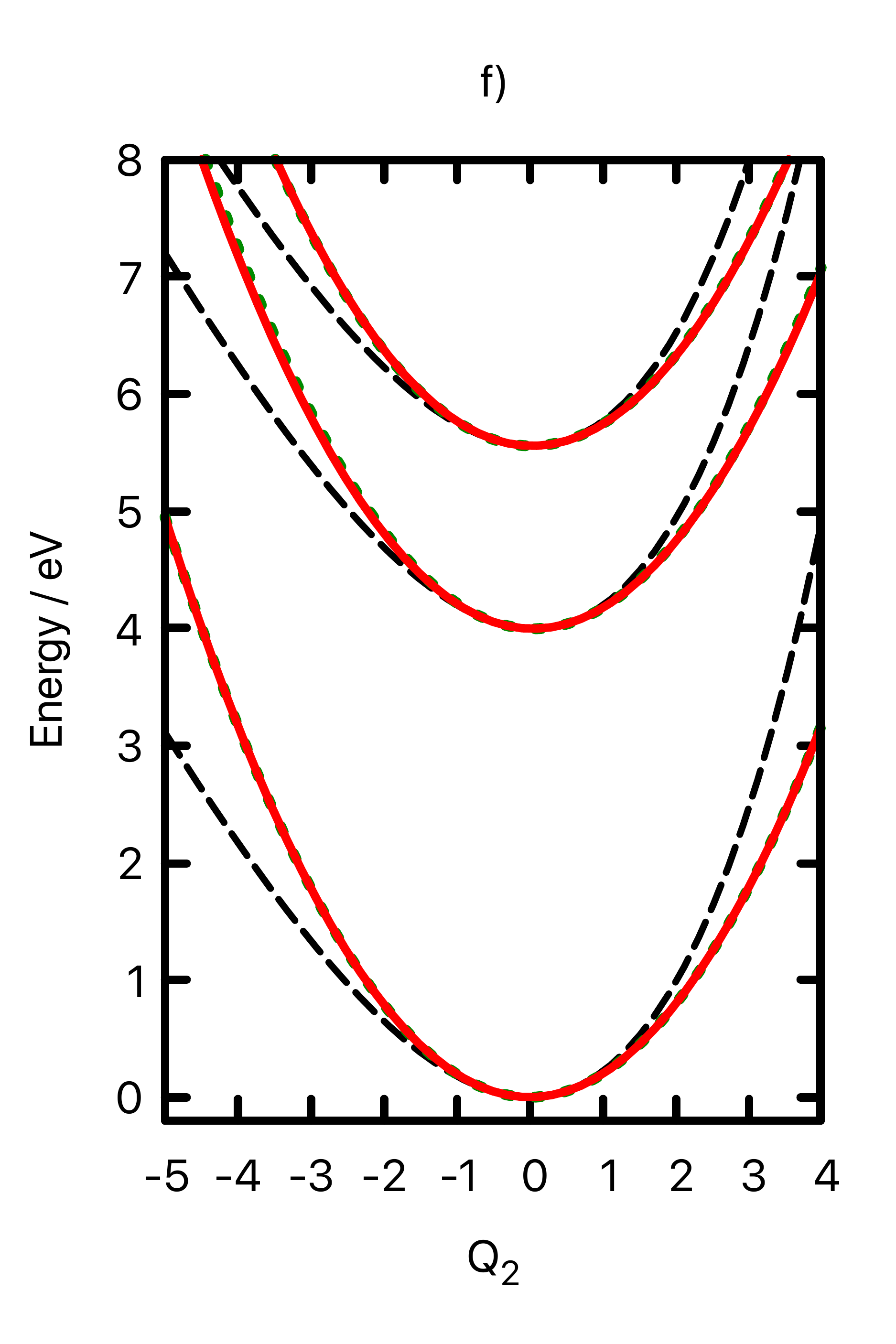"}
    \label{fig:pyrazine_PES_5}
    \end{subfigure}
    \caption{Potential energy curves along the totally symmetric modes and the only mode of B$_{\rm 1g}$ symmetry in pyrazine, ordered by energy. Black, dashed lines: reference; green, dotted lines: displaced and shifted harmonic model; red, solid lines: BO approximation to the linear coupling Hamiltonian; blue, dashed lines in b): one-parameter fit to a harmonic coupling model. Displacements are given in dimensionless coordinates. Energy differences to the ground state minimum are given in eV.}
    \label{fig:pyrazine_PES}
\end{figure}

Figure~\ref{fig:pyrazine_PES} compares PES slices along these modes from our approximate model Hamiltonian and reference calculations with the full electronic Hamiltonian. For comparison, displaced harmonic curves are shown according to
\begin{equation}
    E(Q_\modeone) = \frac{1}{2}\omega_i Q_\modeone^2 + \left.\frac{dE_n}{dQ_\modeone}\right|_{0} + E_n(0) - E_0(0),\label{eq:harmonic_model}
\end{equation}
where $E_n$ is the energy of the $n$-th electronic state and the derivatives of the TDDFT excited state energies at the reference geometry have been obtained numerically. $Q_\modeone$ is the dimensionless displacement along mode $\modeone$ (see below).

It is evident that the model Hamiltonian does lead to PESs in good agreement with the reference in the vicinity of the reference geometry (energetic minimum). At larger displacements, inaccuracies appear as expected, especially in the presence of anharmonicities (Panels c and f of Figure~\ref{fig:pyrazine_PES}), but for the tuning modes the results stay very close to the purely harmonic curves. A striking observation can be made for the coupling mode (Panel b of Figure~\ref{fig:pyrazine_PES}). The model Hamiltonian is indeed able to reproduce features which significantly differ from the harmonic curves and appear as a change in curvature despite the restriction to linear coupling terms and a fixed harmonic force constant. For this case, the results from the proposed linear coupling Hamiltonian are much closer to the reference than to the purely harmonic model with fixed curvature.

Due to symmetry, the displacement along the coupling mode in a real-time propagation calculation will stay zero (unless vibrational energy along the mode is explicitly added). Thus, in the mean-field approximation no population transfer between the different states in question is possible. As will be shown below, a treatment beyond the mean-field approximation will be able to lift this restriction.

\subsection{Extraction of vibronic coupling parameters}

A common approach to study nonadiabatic dynamics is to set up a parameterized model Hamiltonian. Such model Hamiltonians can be further investigated with a number of methods, most prominently with multiconfigurational time-dependent Hartree. Here, we set up and parameterize a vibronic model starting from our approximate molecular Hamiltonian in order to compare resulting parameters to published values for such parameters. In this way, we test whether the leading order contributions to nonadiabatic coupling are still contained in the approximated molecular Hamiltonian.

We will consider a linear coupling model  in dimensionless coordinates,
\begin{equation}
    Q_\modeone = \sqrt{\frac {\omega_\modeone \mu_\modeone}{\hbar}} X_\modeone,
\end{equation}
as is common in the literature. Such models are of the form
\begin{equation}
    \mymat{H}_{\rm model} =
    \mymat{H}^{0}_{\rm el} +
    \mymat{H}_{\rm HO} +
    \mymat{H}_{\rm c},
\end{equation}
where the electronic reference Hamiltonian is a diagonal matrix containing the $n$ vertical energy levels of the electronic subsystem,
\begin{equation}
  \mymat{H}^{0}_{\rm el} =
  \left(\begin{array}{ccc}
    E_0 & 0          & 0   \\
    0           & \ddots & 0   \\
    0           & 0          & E_n
  \end{array}\right)
\end{equation}
$\mymat{H}_{\rm HO}$ is the unperturbed harmonic oscillator Hamiltonian summed over all modes,
\begin{equation}
  \mymat{H}_{\rm HO} = \sum_\modeone \frac{\omega_\modeone}{2}\left(-\frac{\partial^2}{\partial Q_\modeone^2} + Q_\modeone^2\right) \mymat{1},
\end{equation}
and the coupling is defined as
\begin{equation}
  \mymat{H}_{\rm c} =
  \sum_\modeone{\left(\begin{array}{ccc}
    \kappa^{0}_{\modeone} & \lambda^{0, 1}_{\modeone} & \ddots \\
    \lambda^{0, 1}_{\modeone} & \ddots & \lambda^{n-1, n}_{\modeone}   \\
    \ddots & \lambda^{n-1, n}_{\modeone}
    & \kappa^{n}_{\modeone}
  \end{array}\right) Q_\modeone}.
\end{equation}

The vibronic coupling parameters $\kappa$ and $\lambda$ are usually calculated from excited state Hessians or by fitting to PESs.\cite{Raab1999} However, they can alternatively be obtained from differentiation of the electronic Hamiltonian. Recently, a one-shot strategy for direct calculation of these derivatives has been proposed,\cite{Plasser2019} in which only calculations on a single molecular geometry are required. In this approach, derivatives of the wavefunctions of the different states must be calculated. This restricts the approach to wavefunction methods, which can only be applied to systems of limited size (due to the computational complexity of most such methods) or have a very limited accuracy (in case of HF/CIS).

In contrast to the above methods, which work in a many-particle picture, the approach presented in this work allows for staying in a single-particle picture. The matrix elements of the first derivative of the electronic Hamiltonian in the basis of states $\set{m}$ with respect to nuclear displacements in our framework is
   $ \braket{m|\mathbf{A}_\modeone + c_\modeone|n}$.
Thus, the coupling parameters may be calculated as
\begin{eqnarray}
    \kappa_m &= &{\rm tr}(\mymat{A}_\modeone^\dagger (\mymat{D}_{\rm el}^m - \mymat{D}_{\rm el}^0)) \\
    \lambda_{m, n} &= &{\rm tr}(\mymat{A}_\modeone^\dagger \mymat{D}_{\rm el}^{m, n}),
\end{eqnarray}
where $\mymat{D}_{\rm el}^m$ and $\mymat{D}_{\rm el}^{m,n}$ are the electronic density matrix of state $m$ and the transition density between electronic states $m$ and $n$, respectively. The density difference in the equation for the diagonal elements $\kappa_m$ results from the definition of $c_\modeone$. Both the density difference between ground and excited states and the transition density between excited states can be obtained from Eq.~(56) and Eq.~(57) in Ref.~\citenum{Furche2001}. A full calculation of these requires solving one set of CP-SCF equations for each parameter, which can become the time-dominating step in the overall procedure if many excited states are considered. In our current calculation setup, we neglect the expensive orbital relaxation terms entirely, so that the computational cost of calculating all coupling parameters for all states is dominated by a single LR-TDDFT calculation.

In the above, couplings due to the nuclear kinetic energy operator have been neglected. These would lead to additional terms and additional parameters. For pyrazine, the most relevant of these can be expected to be the zeroth order term coupling the $\rm B_{3u}$ and $\rm B_{2u}$ states along $\nu_{\rm 10a}$, namely the term $-(\hbar^2/\mu_{\rm 10a}) \boldsymbol{\pi}\mymat{\nabla}$, which leads to a parameter (analogous to the above) of
\begin{equation}
  -\frac{\hbar^2}{\mu_{\rm 10a}}{\rm tr}(\mymat{D}_{\rm el}^{\rm B_{3u}, B_{2u}} \boldsymbol{\pi}_{\rm 10a}) = 2.3 \times 10^{-4}\;{\rm eV}.
\end{equation}
Due to the scaling by inverse mass, this parameter is several orders of magnitude smaller than the values for $\kappa$ and $\lambda$. Thus, for this term to become relevant, the system would require a huge momentum without experiencing displacements of the same magnitude, which is not what one would expect in a well-behaved propagation of the system.

\begin{figure}
    \centering
    \includegraphics[width=0.7\textwidth]{"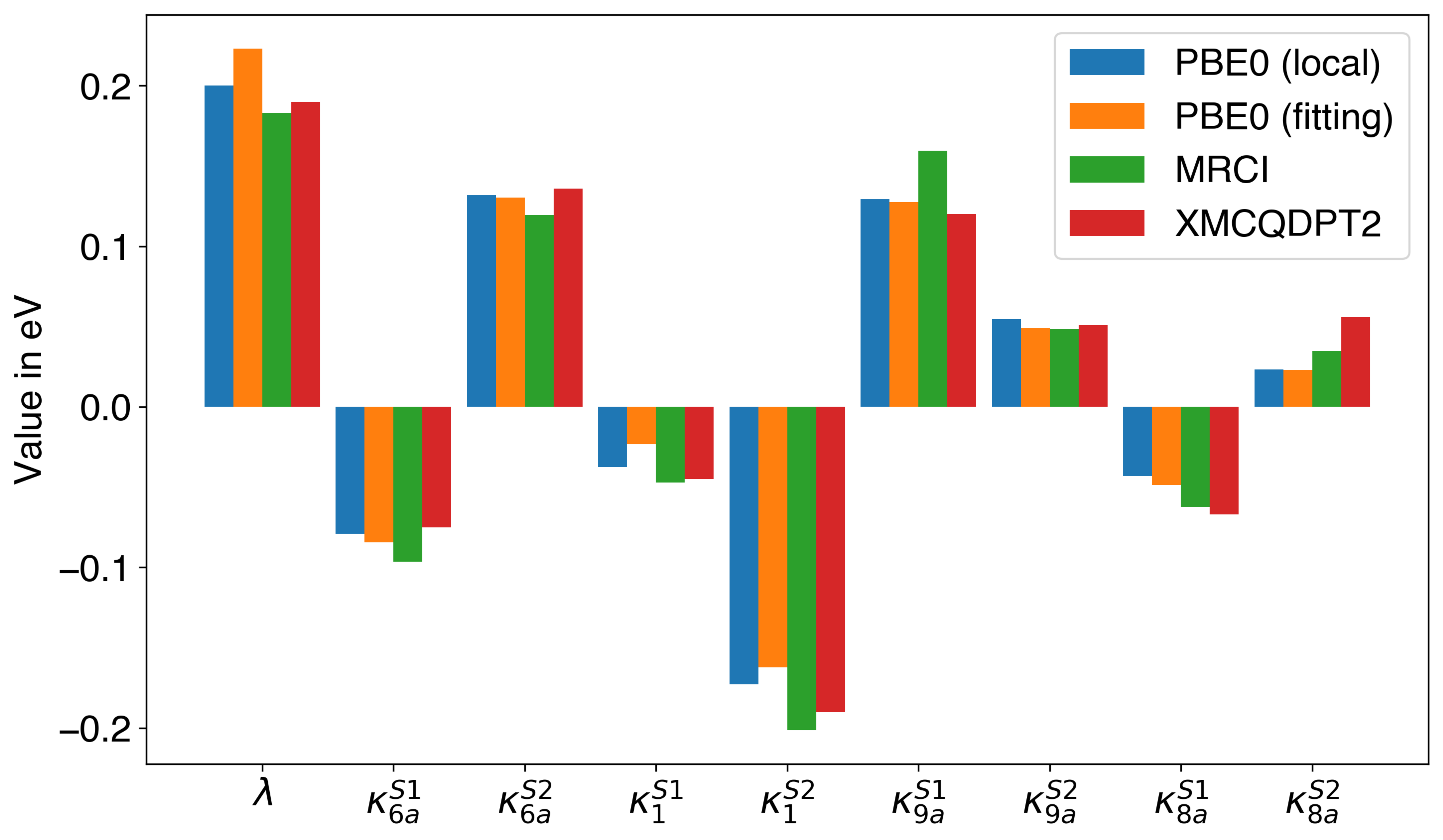"}
    \caption{Selected vibronic coupling parameters presented in Table~\ref{tab:parameters_pyrazine}. Here, S1 is the $^1$B$_{3\mathrm{u}}$ state and S2 is the $^1$B$_{2\mathrm{u}}$ state.}
    \label{fig:parameters_pyrazine}
\end{figure}

Table~\ref{tab:parameters_pyrazine} and Figure~\ref{fig:parameters_pyrazine} compare the most important parameters for a model of the pyrazine system obtained by the approach presented above with results available in the literature. A common procedure to obtain vibronic coupling parameters is to generate PESs and then fit parameters in a model Hamiltonian to the surfaces. For comparison, this procedure has been followed in conjunction with the used mean-field method. The results are shown in Table~\ref{tab:parameters_pyrazine} and Figure~\ref{fig:parameters_pyrazine}. The coupling parameter $\lambda$ was obtained by a simultaneous least-squares fit of both of the eigenvalues of the two-state Hamiltonian
\begin{equation}
    \mymat{H}^{\rm 1p} = \left(\begin{array}{cc}
        \frac{1}{2}\omega_{\rm 10a} Q_{\rm 10a}^2 + E_{\rm B_{3\mathrm{u}}}(0) - E_0(0) & \lambda Q_{\rm 10a} \\
        \lambda Q_{\rm 10a} & \frac{1}{2}\omega_{\rm 10a} Q_{\rm 10a}^2 + E_{\rm B_{2\mathrm{u}}}(0) - E_0(0),
    \end{array}\right)
\end{equation}
in the interval of $Q_{10a}\in[-1,1]$ to the original potential energy curves of the $B_{3\mathrm{u}}$ and $B_{3\mathrm{u}}$ states. The result of this fit is further shown in Panel b of Fig.~\ref{fig:pyrazine_PES} (blue, dashed lines). In addition to this coupling mode, the totally symmetric modes (tuning modes) are usually included in such a model as well. The $\kappa$ parameters are derivatives of the excited state energies. These have been calculated numerically and have already been used to produce the green, dotted curves in Fig.~\ref{fig:pyrazine_PES} (see also Eq.~\ref{eq:harmonic_model}).

\begin{table}
    \centering
    \caption{Vibronic coupling parameter ($\lambda$) coupling the lowest $^1$B$_{3\mathrm{u}}$ state to the lowest $^1$B$_{2\mathrm{u}}$ state and electron-vibrational coupling constants ($\kappa$) for the totally symmetric (A$_\mathrm{g}$) modes for the lowest three singlet excited states of pyrazine. Results obtained from local calculations as described in the text are compared to values reported in the literature. All numbers are given in $\mathrm{eV}$. }
    \begin{tabular}{lccccccc}
    \hline\hline
        &&\hspace*{10pt}& \multicolumn{1}{c}{Local} &\hspace*{10pt}& \multicolumn{3}{c}{Fitting to PESs} \\
        \cline{4-4}\cline{6-8}
        & Mode  && PBE0 && PBE0 
        & MRCI\cite{Woywod1994} &  XMCQDPT2\cite{Sala2015} \\ \hline
        $\lambda$ & $\nu_{10a}$ &&  $\phantom{+}0.200\phantom0$ && $\phantom{+}0.223\phantom0$ & $\phantom{+}0.183\phantom0$ & $\phantom{+}0.190$ \\[2ex]
        $^1$B$_{3\mathrm{u}}$($\mathrm{n}\pi^*$)
                       & $\nu_{6a}$ && $-0.0791$ && $-0.0844$ & $-0.0964$ & $-0.075$ \\
                       & $\nu_{1}$ && $-0.0376$ && $-0.0231$ & $-0.0470$ & $-0.045$  \\
                       & $\nu_{9a}$ && $\phantom{+}0.1295$ && $\phantom{+}0.1275$ & $\phantom{+}0.1594$ & $\phantom{+}0.120$ \\
                       & $\nu_{8a}$ && $-0.0432$ && $-0.0488$ & $-0.0623$ & $-0.067$ \\
                       & $\nu_{2}$ && $-0.0175$ && $-0.0219$ & $\phantom{+}0.0368$ & \\[2ex]
        $^1$A$_{\mathrm{u}}$($\mathrm{n}\pi^*$)
                       & $\nu_{6a}$ && $-0.1709$ && $-0.1751$ & $ $ & $-0.162$ \\
                       & $\nu_{1}$ && $-0.1076$ && $-0.0935$ & $ $ & $-0.088$  \\
                       & $\nu_{9a}$ && $-0.0563$ && $-0.0607$ & $ $ & $-0.064$ \\
                       & $\nu_{8a}$ && $-0.4546$ && $-0.4524$ & $ $ & $-0.413$ \\
                       & $\nu_{2}$ && $-0.0716$ && $-0.0742$ & $ $ & $ $\\[2ex] 
        $^1$B$_{2\mathrm{u}}$($\pi\pi^*$)
                       & $\nu_{6a}$ && $\phantom{+}0.1318$ && $\phantom{+}0.1302$ & $\phantom{+}0.1193$ & $\phantom{+}0.136$ \\
                       & $\nu_{1}$ && $-0.1728$ && $-0.1621$ & $-0.2012$ & $-0.190$ \\
                       & $\nu_{9a}$ && $\phantom{+}0.0547$ && $\phantom{+}0.0489$ & $\phantom{+}0.0484$ & $\phantom{+}0.051$ \\
                       & $\nu_{8a}$ && $\phantom{+}0.0234$ && $\phantom{+}0.0230$ & $\phantom{+}0.0348$ & $\phantom{+}0.056$ \\
                       & $\nu_{2}$ && $-0.0138$ && $-0.0110$ & $\phantom{+}0.0211$ &  \\ 
                       \hline\hline
    \end{tabular}
    \label{tab:parameters_pyrazine}
\end{table}

The results obtained from single-shot calculations on the proposed Hamiltonian compare very well with those obtained from fitting to PESs and are compatible with previously published parameters for pyrazine. The different sign obtained for the $\nu_2$ mode clearly results from the used mean-field method. Some of the remaining descripancies for modes $\nu_1$ and $\nu_2$ can be attributed to anharmonicities. Additional results for different computational settings (functionals, basis sets, integration grids) can be found in Table~S2 in the supplementary material.

As a conclusion, the obtained vibronic coupling parameters lead to a vibronic model Hamiltonian similar to models used earlier to study the excited state dynamics of pyrazine. Thus, similar results for properties like rate constants can be obtained from the presented parameterization strategy.

More importantly, it supplies evidence that the model electron-vibration Hamiltonian proposed in this work does contain the key features for studying nonadiabatic effects.

\subsection{Comparison with other choices for the electronic basis}
In order to compare the performance of the choice $\mymat{T}\ofX = \mymat{S}^{-\frac{1}{2}}\ofX$ with the canonical MO basis ($\mymat{T}\ofX = \mymat{C}\ofX$) and the basis suggested by Troisi ($\mymat{T}\ofX = \mymat{C}\ofzero - \mymat{S}^{-1}\ofzero\mymat{S}^{(0,1)}\ofzero\mymat{C}\ofzero\myvec{X}$), we numerically calculate the slopes and curvatures of the TDDFT excited states of the approximate Hamiltonians resulting from the different basis choices. Hereby, we concentrate on the most essential features of the PESs of pyrazine, which are the slopes of the lowest $^1$B$_{3\mathrm{u}}$ and $^1$B$_{2\mathrm{u}}$ excited states along the most important tuning mode $\nu_{6a}$  as well as the change of the curvature of these states (with respect to the ground state curvature) along this mode and the coupling mode $\nu_{10a}$. The results are collected in Table~\ref{tab:slopes_and_curvatures}.

\begin{table}
    \centering
    \caption{Slopes and changes in curvature (in eV) of the potential energy curves of the $^1{\rm B}_{3\mathrm{u}}$ and $^1{\rm B}_{2\mathrm{u}}$ states of pyrazine along the most important tuning mode ($\nu_{6a}$) and the coupling mode ($\nu_{10a}$) obtained from the linear coupling Hamiltonian using different electronic bases. Symmetric refers to the symmetrically orthgonalized model. Reference results are obtained from the potential energy curves of the original electronic Hamiltonian.}
    \begin{tabular}{llcccc}
\hline\hline
&& Symmetric & Canonical & Troisi & Reference \\\hline
$^1{\rm B}_{3\mathrm{u}}$  \hspace*{15pt}
 &       $\partial E/\partial Q_{6a}$        & $-0.078$ & $-0.084$ & $-0.104$ & ${-0.084}$ \\
 &
        $\Delta\partial^2 E/\partial Q_{6a}^2$ & $-0.001$ & $\phantom{+}0.0\phantom{00}$ & $-0.012$ & ${\phantom{+}0.0\phantom{00}}$\\
 &
        $\Delta\partial^2 E/\partial Q_{10a}^2$ & $-0.065$ & $\phantom{+}0.0\phantom{00}$ & $-0.180$ & ${-0.092}$ \\[2ex]
$^1{\rm B}_{2\mathrm{u}}$ &        
        $\partial E/\partial Q_{6a}$            & $\phantom{+}0.120$ & $\phantom{+}0.130$ & $\phantom{+}0.107$ & ${\phantom{+}0.130}$ \\
 &
        $\Delta\partial^2 E/\partial Q_{6a}^2$ & $-0.004$& $\phantom{+}0.0\phantom{00}$ & $-0.003$ & ${-0.021}$\\
 &        
        $\Delta\partial^2 E/\partial Q_{10a}^2$ & $\phantom{+}0.017$& $\phantom{+}0.0\phantom{00}$ & $\phantom{+}0.057$ & ${\phantom{+}0.038}$\\
    \hline\hline
    \end{tabular}
    \label{tab:slopes_and_curvatures}
\end{table}

For the canonical MO basis, the slopes match those of the original PESs and the change in curvature is zero, both by construction. Thus, the significant curvature changes along $\nu_{10a}$ are not captured. The symmetrically orthogonalised basis we suggest for usage with our model does correctly capture significant amounts of the curvature change at the expense of small errors in the slopes. The Troisi basis tends to too large absolute values. For the $^1{\rm B}_{2\mathrm{u}}$ state, this results in a negative curvature of the potential energy curve along $\nu_{10a}$ (the ground state curvature amounts to $0.118\;{\rm eV}$, thus the curvature of the $^1{\rm B}_{2\mathrm{u}}$ state becomes $-0.072\;{\rm eV}$). Such a negative curvature leads to unbound states, which can lead to severe problems in practical calculations. In fact, we did observe severe errors at large displacements when using the Troisi basis in preliminary calculations, which are probably caused by mixing in of excited states with negative curvatures.

Overall, the symmetrically orthogonalised basis shows the best performance of the tested bases for the level of approximation used in this work.

\subsection{Coupled Time-Dependent Linear Response}
In coupled linear response calculations, both the hermitian and antihermitian terms resulting from the nuclear kinetic energy operator can in principle lead to a contribution.

In practice, the results for pyrazine from coupled time-dependent linear response calculations show only tiny deviations from separate electronic and vibrational excitations. The electronic excitations are vertical. The linear response framework is only able to calculate single excitations. Alas, excitations from the overall ground state to the (mostly) vibrational ground state of a (mostly) excited electronic state formally represents a multiply excited state when the minimum of the excited state surface shows a significant displacement. Thus, these excitations cannot be captured in the linear response framework. Also, the linear response framework is not able to capture all relevant correlation effects.

\subsection{A pilot study on the role of correlations}
The above findings show that the inclusion of correlation is crucial for observing key nonadiabatic effects in problems such as the photophysics of pyrazine. Here, the correlation between electrons and vibrations is of particular importance. While correlated calculations using the model Hamiltonian will be the subject of a future publication, we here demonstrate that even the simplest correlated treatment --- a strictly limited configuration interaction calculation supplemented with key electron-vibration double excitations --- produces the key nonadiabatic phenomenon of population transfer between the S$_2$ and S$_1$ states. In particular we included excitations from the electronic HOMO or HOMO-1 to the electronic LUMO (the key elements of the $^1$B$_{3\mathrm{u}}$ and $^1$B$_{2\mathrm{u}}$ excitations) together with a vibrational excitation along $\nu_{10a}$.

This truncated CISD calculation has dimension $1 + n^{\rm el}_{\rm occ} n^{\rm el}_{\rm virt} + 6\ n^{\rm vib}_{\rm occ} n^{\rm vib}_{\rm virt} + 2 = 3978$,
with successive terms arising from the ground state, the singly excited electronic space, the singly excited vibrational space for the coupling mode and the five tuning modes, and two mixed doubly excited states. The vibrational singles consist of just one excitation per mode ($n_{\rm vib}^{\rm virt} = 1$), because of the linear coupling in this model.

Hartree--Fock calculations result in significantly different orbital energies and also change the order of the $^1$A$_{\mathrm{u}}$ and $^1$B$_{2\mathrm{u}}$ states, so we pragmatically used Kohn--Sham orbitals and the Tamm--Dancoff matrix instead of Hartree--Fock orbitals and the CIS matrix for the singles block of the calculation.

In comparison with TDDFT results in the Tamm--Dancoff approximation, the excitation energies of the $^1$B$_{3\mathrm{u}}$ and $^1$B$_{2\mathrm{u}}$ states are shifted lower by a small amount, $0.01\;\mathrm{eV}$ and $0.07\;\mathrm{eV}$, respectively.
We performed time propagation using this limited CISD Hamiltonian, starting in the product state composed of the $^1$B$_{2\mathrm{u}}$ electronic and vibrational ground state.

The $^1$B$_{2\mathrm{u}}$ population oscillates between 1 and around 0.3 with an oscillation period of $14\;\mathrm{fs}$, and with practically all population transfer to the $^1$B$_{3\mathrm{u}}$ state. The computed period is remarkably close to the experimental lifetime of the $^1$B$_{2\mathrm{u}}$ state of $22\;\pm\;3\;\mathrm{fs}$\cite{Suzuki2010} given the crude approximations we applied in this exploratory correlated calculation.
In line with the findings of a similar work using a many-particle picture,\cite{DurgaPrasad1992} dephasing cannot be observed in this calculation, because there are no couplings between the electronic excitation and the vibrational tuning modes. Nevertheless, it provides a further indication that the model Hamiltonian captures the key phenomenology, and that correlated (e.g. coupled-cluster) theories based on this Hamiltonian should provide a rich alternative avenue for exploring nonadiabatic dynamics, without reference to potential energy surfaces, conical intersections, or diabatization.

\section{Conclusions}
We have derived an electron-vibration model Hamiltonian that contains only up to two-body terms and can be constructed for any molecular system for which a ground-state Hessian can be computed at a reference geometry. The model follows the spirit of methods to describe electron-phonon coupling in condensed-matter physics, but uses atom-centred non-orthogonal single-particle basis functions. For pyrazine the choice to define position-dependence in terms of the symmetrically orthogonalised basis leads to a model that captures key effects of the full molecular system using only linear coupling terms. 

While the method we are proposing does not involve calculation of PESs, they have nevertheless proven to be a helpful way to assess the accuracy of the model. PESs extracted from our model Hamiltonian closely resemble those of the full molecular Hamiltonian in the vicinity of the reference geometry, and match them better than a simple harmonic fit. Qualitatively correct vibronic coupling parameters can be extracted at essentially the cost of a nuclear Hessian plus linear-response TDDFT calculation. Throughout the calculation of these parameters, no PESs need to be calculated and no diabatization is necessary.

This work has established a model Hamiltonian for molecular nonadiabatic effects. The next task is to elaborate the full range of wavefunction-based quantum chemistry methods for this Hamiltonian, building an alternative framework for studying nonadiabatic effects. Preliminary steps in that direction include coupled electron-vibration mean-field theory, and linear response theory. As expected, neither lead to significant vibronic effects, because the key phenomenon can be regarded as a ``double" that couples simultaneous electronic and vibrational excitations. 
As a proof of principle, we have shown that a correlated propagation containing only the most relevant coupled electron-vibration excitations shows qualitatively correct population transfer from the S$_2$ to the S$_1$ state of pyrazine. Such effects would be fully captured in a correlated framework such as coupled-cluster theory, provided the cluster operator includes these double excitations. 

Our broader aim is to construct a systematically improvable hierarchy of quantum-chemistry-like methods for studying nonadiabatic effects. The Hamiltonian can be systematically improved by including higher-order terms in the Taylor expansions that underpin the derivation, and by removing the Fock approximation of coupling to 2-electron terms. While some extensions can be achieved while remaining in the framework of a two-body Hamiltonian, typically these additional effects are described by three-body or higher order terms. The wavefunction can be systematically improved in a coupled-cluster framework, where extension to Hamiltonians that include both electrons and other degrees of freedom is already a proven technology.\cite{Mordovina2020,White2020}
Importantly, all of this can be achieved without reference to potential energy surfaces, conical intersections, or diabatization. 
The combination of these two ideas --- systematically improvable model Hamiltonians and a systematically improvable framework for describing their quantum states and dynamics --- provides a roadmap for the development of a powerful new family of polynomial scaling theories for nonadiabatic dynamics.

\begin{acknowledgments}
 We are grateful for funding from the Engineering and Physical Sciences Research Council (EPSRC) through grants EP/R014493/1 and EP/R014183/1. One of us (C.B.A.B.) is funded through the EPSRC Centre for Doctoral Training in Theory and Modelling in Chemical Sciences (EP/L015722/1).
 We gratefully acknowledge Prof.\@ Garnet Chan and Dr.\@ Marat Sibaev for helpful discussions.
\end{acknowledgments}

The data that supports the findings of this study are available within the article and its supplementary material.

One of the authors (F.R.M.) is co-founder and CTO of Entos Inc. The other authors declare no conflict of interest. 

\bibliography{manuscript}

\begin{thebibliography}{6}
\bibitem [{a()}]{a}%
  \BibitemOpen
  \bibinfo {note} {Normal modes and frequencies from the calculation with B3LYP
  have been used.}\BibitemShut {Stop}%
\end{thebibliography}

\end{document}


\title{Supplementary Material --- Coupling electrons and vibrations in molecular quantum chemistry}

\author{Thomas Dresselhaus}
\author{Callum B. A. Bungey}
\affiliation{Centre for Computational Chemistry, School of Chemistry, University of Bristol, Bristol BS8 1TS, United Kingdom}
\author{Peter J. Knowles}
\affiliation{School of Chemistry, Cardiff University, Main Building, Park Place, Cardiff CF10 3AT, United Kingdom}
\author{Frederick R. Manby}
\email{fred.manby@bristol.ac.uk.}
\affiliation{Centre for Computational Chemistry, School of Chemistry, University of Bristol, Bristol BS8 1TS, United Kingdom}

\date{\today}

\maketitle

\section{Anti-Hermiticity of $1 + 2 X_\modeone\nabla_\modeone$}
As derived in the main article, the first-order term containing $\pi_{IJ}^\modeone\ofX$ along $\modeone$ is
\begin{equation}
    \left[\nabla_\modeone\pi^{\modeone}_{IJ}\right](1 + 2 X_\modeone\nabla_\modeone)
\end{equation}
The anti-Hermiticity in vibrational space is not trivial to see for this term and is shown in the following. Standard bosonic creation and annihilation operators are used. In the harmonic oscillator basis,
\begin{align}
    \hat{X}_\modeone &= \sqrt{\frac{\hbar}{2\mu_\modeone \omega_\modeone}}\left(b_\modeone^\dagger + b_\modeone\right)\\
    \nabla_\modeone &= -\sqrt{\frac{\mu_\modeone \omega_\modeone}{2\hbar}}\left(b_\modeone^\dagger - b_\modeone\right),
\end{align}
where $\mu_\modeone$ and $\omega_\modeone$ are the reduced mass and the vibrational frequency of the harmonic oscillator describing mode $\modeone$. Inserting this yields 
\begin{align}
    1 + 2\hat{X}_\modeone\nabla_\modeone &=
    1 - 2 \left(b_\modeone^\dagger b_\modeone^\dagger - b_\modeone^\dagger b_\modeone + b_\modeone b_\modeone^\dagger - b_\modeone b_\modeone \right)\\
    &= -2 \left(b_\modeone^\dagger b_\modeone^\dagger - b_\modeone b_\modeone\right)
\end{align}
for the first quantized vibrational term. The last equality follows from the commutation relations and the anti-Hermiticity is now obvious.

\section{Derivatives of $\boldsymbol{\pi}$ and $\mymat{\Pi}$}
In the main manuscript, definitions of the matrices $\boldsymbol{\pi}_\modeone$ and $\mymat{\Pi}_\modeone$, which couple the electronic and vibrational systems as a result of $\hat{T}^\modeone_{\rm nuc}$, have been given. As we are only considering mode $\modeone$ here, the explicit notation of that index is dropped. The derivatives along mode $\modeone$ are
\begin{align}
{\pi}_{pq}^{(1)} &=
\left(\braket{\phi_p^{(1)}|\phi_q^{(1)}} +
\braket{\phi_p|\phi_q^{(2)}}\right)\\
{\Pi}_{pq}^{(1)} &=
\left(\braket{\phi_p^{(2)}|\phi_q^{(1)}} + \braket{\phi_p^{(1)}|\phi_q^{(2)}}\right)
,
\end{align}
where
\begin{align}
\braket{\phi_p^{(1)}|\phi_q^{(1)}} &=  \Pi^\modeone_{pq}\\
\braket{\phi_p|\phi_q^{(2)}} &=\left[
\mymat{T}^\dagger\mymat{S}^{(0, 2)}\mymat{T} +
2 \mymat{T}^\dagger \mymat{S}^{(0, 1)}\mymat{T}^{(1)} +
\mymat{T}^\dagger\mymat{S}\mymat{T}^{(2)}\right]_{pq}\\
\braket{\phi_p^{(2)}|\phi_q^{(1)}} &=
\left[\mymat{T}^{(2)\dagger}\mymat S^{(0, 1)}\mymat{T} +
\mymat{T}^{(2)\dagger}\mymat S\mymat{T}^{(1)} +
2 \mymat{T}^{(1)\dagger}\mymat S^{(1, 1)}\mymat{T}\right.\\
&+\left.
\mymat{T}^{(1)\dagger}\mymat{S}^{(1, 0)}\mymat{T}^{(1)} +
\mymat{T}^\dagger\mymat S^{(2, 1)}\mymat{T} +
\mymat{T}^\dagger\mymat S^{(2, 0)}\mymat{T}^{(1)}\right]_{pq}\nonumber\\
\braket{\phi_p^{(1)}|\phi_q^{(2)}} &=\left[
\mymat{T}^{(1)\dagger}\mymat{S}^{(0, 2)}\mymat{T} +
\mymat{T}^{(1)\dagger}\mymat S^{(0, 1)}\mymat{T}^{(1)} +
\mymat{T}^{(1)\dagger}\mymat{S}\mymat{T}^{(2)} \right.\\ &+ \left.
\mymat{T}^\dagger\mymat S^{(1, 2)}\mymat{T} +
2 \mymat{T}^\dagger\mymat{S}^{(1,1)}\mymat{T}^{(1)} +
\mymat{T}^\dagger\mymat S^{(1,0)}\mymat{T}^{(2)}\right]_{pq}\nonumber
\end{align}

\section{Derivatives of $\mymat{T}\ofX = \mymat{S}^{-\frac{1}{2}}\ofX$}
The derivatives of $\mymat{T} = \mymat{S}^{-\frac{1}{2}}$ can be calculated straightforwardly and separately for each mode. Given the identity $\mymat{T}\mymat{S}\mymat{T} = \mymat{1}$ the gradient is
\begin{equation}
    \mymat{T}'\mymat{S}\mymat{T} +
    \mymat{T}\mymat{S}'\mymat{T} +
    \mymat{T}\mymat{S}\mymat{T}' = \mymat{0},
\end{equation}
where the prime denotes a partial derivative with respect to mode $\modeone$.
This is a Sylvester equation for the unknown $\mymat{T}'$, for which a solver is implemented e.g.\ in the linear algebra library {\sc armadillo}.\cite{armadillo} Further derivatives can be obtained by further differentiating the above equation and require only further derivatives of the overlap matrix and the results from the previous derivatives. E.g., the second derivative is obtained by forming the derivative of the above equation:
\begin{align}
    \mymat{T}'\mymat{S}\mymat{T} +
    \mymat{T}\mymat{S}'\mymat{T} &+
    \mymat{T}\mymat{S}\mymat{T}' = \mymat{0}\nonumber \\
    \Rightarrow
    \mymat{T}''\mymat{S}\mymat{T} +
    \left(
    2 \mymat{T}'\mymat{S}'\mymat{T} +
    2 \mymat{T}'\mymat{S}\mymat{T}' \right.&\left.+
    2\mymat{T}\mymat{S}'\mymat{T}' +
    \mymat{T}\mymat{S}''\mymat{T}
    \right) +
    \mymat{T}\mymat{S}\mymat{T}'' = \mymat{0}
\end{align}
Note that given $\mymat{T}'$ and $\mymat{S}''$ this again results in a Sylvester equation, where all terms which are independent of $\mymat{T}''$ can be collected into a single matrix as indicated by the parentheses. The same strategy can be iteratively applied in order to obtain higher-order derivatives, or also mixed derivatives w.r.t.\@ the coordinates of more than one mode.

\section{Vibronic coupling parameters for pyrazine}
The following table denotes results in analogy to Table~I of the main manuscript obtained with varied computational settings. The first column of results is the one presented in the main manuscript. The results of all other columns differ only in the listed functional/basis set/integration grid. DZ and QZ are double-$\zeta$ and quadruple-$\zeta$ basis sets,\cite{Weigend2008} and the ``large grid'' calculation uses an integration grid with $115316$ grid points in comparison to $64360$ grid points used in all other calculations.
\\[\intextsep]
  \begin{minipage}{\linewidth}
    \centering%
    \label{tab:vibronic coupling parameters}
    \tabcaption{Vibronic coupling parameter ($\lambda$) coupling the lowest $^1$B$_{3\mathrm{u}}$ state to the lowest $^1$B$_{2\mathrm{u}}$ state and electron-vibrational coupling constants for the totally symmetric (A$_\mathrm{g}$) modes for the lowest three singlet excited states of pyrazine. Different computational settings are compared to the setup used in the main manuscript (column ``PBE0''), see the text for details. All numbers are given in $\mathrm{eV}$.}
    \begin{tabular}{lcccccccc}\hline\hline
        & Mode & \hspace{2.5ex}PBE0\hspace{2.5ex}  & B3LYP\cite{Stephens1994} & CAMB3LYP\footnote{Normal modes and frequencies from the calculation with B3LYP have been used.}$^{,}$\cite{Yanai2004} & PBE\cite{Perdew1996} & DZ & QZ & large grid \\ \hline
        $\lambda$ & $\nu_{10a}$ & $\phantom{+}0.2003$ & $\phantom{+}0.1964$ & $\phantom{+}0.2012$ & $\phantom{+}0.1951$ & $\phantom{+}0.2048$ & $\phantom{+}0.1996$ & $\phantom{+}0.2002$ \\[2ex]
        $^1$B$_{3\mathrm{u}}$($\mathrm{n}\pi^*$)
                       & $\nu_{1}$ & $-0.0376$ & $-0.0422$ & $-0.0567$ & $-0.0138$ & $-0.0283$ & $-0.0370$ & $-0.0375$\\
                       & $\nu_{2}$ & $-0.0175$ & $-0.0204$ & $-0.0164$ & $-0.0270$ & $-0.0148$ & $-0.0166$ & $-0.0175$\\
                       & $\nu_{6a}$ & $-0.0791$ & $-0.0764$ & $-0.0814$ & $-0.0750$ & $-0.0818$ & $-0.0790$ & $-0.0791$\\
                       & $\nu_{8a}$ & $-0.0432$ & $-0.0411$ & $-0.0444$ & $-0.0318$ & $-0.0227$ & $-0.0470$ & $-0.0432$\\
                       & $\nu_{9a}$ & $\phantom{+}0.1295$ & $\phantom{+}0.1237$ & $\phantom{+}0.1306$ & $\phantom{+}0.1229$ & $\phantom{+}0.1306$ & $\phantom{+}0.1298$ & $\phantom{+}0.1219$\\[2ex]
        $^1$A$_{\mathrm{u}}$($\mathrm{n}\pi^*$)
                       & $\nu_{1}$ & $-0.1076$ & $-0.1117$ & $-0.1237$ & $-0.0919$ & $-0.1002$ & $-0.1081$ & $-0.1077$\\
                       & $\nu_{2}$ & $-0.0716$ & $-0.0742$ & $-0.0695$ & $-0.0827$ & $-0.0737$ & $-0.0683$ & $-0.0715$\\
                       & $\nu_{6a}$ & $-0.1709$ & $-0.1665$ & $-0.1719$ & $-0.1681$ & $-0.1790$ & $-0.1692$ & $-0.1710$\\
                       & $\nu_{8a}$ & $-0.4546$ & $-0.4418$ & $-0.4732$ & $-0.4077$ & $-0.4513$ & $-0.4537$ & $-0.4548$\\
                       & $\nu_{9a}$ & $-0.0563$ & $-0.0720$ & $-0.0748$ & $-0.0300$ & $-0.0169$ & $-0.060$ & $-0.0559$\\[2ex]
        $^1$B$_{2\mathrm{u}}$($\pi\pi^*$)
                       & $\nu_{1}$ & $-0.1728$ & $-0.1694$ & $-0.1852$ & $-0.1551$ & $-0.1732$ & $-0.1701$ & $-0.1727$\\
                       & $\nu_{2}$ & $-0.0138$ & $-0.0145$ & $-0.0169$ & $-0.0116$ & $-0.0128$ & $-0.0126$ & $-0.0139$\\
                       & $\nu_{6a}$ & $\phantom{+}0.1318$ & $\phantom{+}0.1277$ & $\phantom{+}0.1345$ & $\phantom{+}0.1194$ & $\phantom{+}0.1230$ & $\phantom{+}0.1328$ & $\phantom{+}0.1319$\\
                       & $\nu_{8a}$ & $\phantom{+}0.0234$ & $\phantom{+}0.0231$ & $\phantom{+}0.0199$ & $\phantom{+}0.0257$ & $\phantom{+}0.0300$ & $\phantom{+}0.0173$ & $\phantom{+}0.0234$\\
                       & $\nu_{9a}$ & $\phantom{+}0.0547$ & $\phantom{+}0.0527$ & $\phantom{+}0.0556$ & $\phantom{+}0.0577$ & $\phantom{+}0.0644$ & $\phantom{+}0.053$ & $\phantom{+}0.0546$\\ \hline
    \end{tabular}
  \end{minipage}
\\[\intextsep]

\newpage
\bibliography{SI}